\documentclass[]{pasj01}

\Received{}
\Accepted{}
 
\usepackage{url}  
\usepackage{graphicx}  
\usepackage{dcolumn} 
\usepackage{natbib} 
\usepackage{footnote}

\makeatletter 
\newcommand{\figcaption}[1]{\def\@captype{figure}\caption{#1}}
\newcommand{\tblcaption}[1]{\def\@captype{table}\caption{#1}}
\makeatother

\begin{document} 

\title{ 
Subaru High-z Exploration of Low-Luminosity Quasars (SHELLQs). IX. Identification of Two Red Quasars at z $>$ 5.6
  }

\author{Nanako Kato\altaffilmark{1}}%
\email{kato@cosmos.phys.sci.ehime-u.ac.jp}

\author{Yoshiki Matsuoka\altaffilmark{2}}
\email{yk.matsuoka@cosmos.ehime-u.ac.jp}

\author{Masafusa Onoue\altaffilmark{3}}
\author{Shuhei Koyama\altaffilmark{2}}
\author{Yoshiki Toba\altaffilmark{4, 5, 2}}
\author{Masayuki Akiyama\altaffilmark{6}}
\author{Seiji Fujimoto\altaffilmark{7, 8}}
\author{Masatoshi Imanishi\altaffilmark{9, 10}}
\author{Kazushi Iwasawa\altaffilmark{11}}
\author{Takuma Izumi\altaffilmark{9, 10}}
\author{Nobunari Kashikawa\altaffilmark{12}}
\author{Toshihiro Kawaguchi\altaffilmark{13}}
\author{Chien-Hsiu Lee\altaffilmark{14}}
\author{Takeo Minezaki\altaffilmark{15}}
\author{Tohru Nagao\altaffilmark{2}}
\author{Akatoki Noboriguchi\altaffilmark{1}}
\author{Michael A. Strauss\altaffilmark{16}}

\altaffiltext{1}{Graduate School of Science and Engineering, Ehime University, Matsuyama, Ehime 790-8577, Japan.}
\altaffiltext{2}{Research Center for Space and Cosmic Evolution, Ehime University, Matsuyama, Ehime 790-8577, Japan.}
\altaffiltext{3}{Max Planck Institut f\"ur Astronomie, K\"onigstuhl 17, D-69117 Heidelberg, Germany}
\altaffiltext{4}{Department of Astronomy, Kyoto University, Kitashirakawa-Oiwake-cho, Sakyo-ku, Kyoto 606-8502, Japan}
\altaffiltext{6}{Academia Sinica Institute of Astronomy and Astrophysics, 11F of Astronomy-Mathematics Building, AS/NTU, No.1, Section 4, Roosevelt Road, Taipei 10617, Taiwan}
\altaffiltext{7}{Cosmic DAWN Center}
\altaffiltext{8}{Niels Bohr Institute, University of Copenhagen, Lyngbyvej2, DK-2100, Copenhagen, Denmark}
\altaffiltext{9}{National Astronomical Observatory of Japan, Mitaka, Tokyo 181-8588, Japan}
\altaffiltext{10}{Department of Astronomical Science, Graduate University for Advanced Studies (SOKENDAI), Mitaka, Tokyo 181-8588, Japan}
\altaffiltext{11}{ICREA and Institut de Ci{\`e}ncies del Cosmos, Universitat de Barcelona, IEEC-UB, Mart{\'i} i Franqu{\`e}s, 1, 08028 Barcelona, Spain.}
\altaffiltext{12}{Department of Astronomy, School of Science, The University of Tokyo, Tokyo 113-0033, Japan}
\altaffiltext{13}{Department of Economics, Management and Information Science, Onomichi City University, Onomichi, Hiroshima 722-8506, Japan}
\altaffiltext{14}{National Optical Astronomy Observatory, 950 North Cherry Avenue, Tucson, AZ 85719, USA}
\altaffiltext{15}{Institute of Astronomy, School of Science, University of Tokyo, Mitaka, Tokyo 181-0015, Japan}
\altaffiltext{16}{Princeton University Observatory, Peyton Hall, Princeton, NJ 08544, USA}


\KeyWords{quasars: general --- galaxies: high-redshift  --- early universe ---}

\maketitle

\begin{abstract}
We present the first discovery of dust-reddened quasars (red quasars) in the high-$z$ universe ($z >5.6$).
This is a result from the Subaru High-$z$ Exploration of Low-Luminosity Quasars (SHELLQs) project, which is based on the sensitive multi-band optical imaging data produced by the Hyper Suprime-Cam (HSC) Subaru Strategic Program survey.
We identified four red quasar candidates from the spectroscopically confirmed 93 high-$z$ quasars in the SHELLQs sample, based on detections in the Wide-field Infrared Survey Explorer ($\it WISE$) data at 3.4 and 4.6 $\mu$m (rest-frame $\sim$5000--6500 \AA).
The amount of dust reddening was estimated with spectral energy distribution (SED) fits over optical and mid-infrared wavelengths.
Two of the four candidates were found to be red quasars with dust reddening of $E(B-V) > 0.1$.
The remaining SHELLQs quasars without individual $\it WISE$ detections are significantly fainter in the $\it WISE$ bands and bluer than the red quasars, although we did detect them in the $W1$ band in a stacked image.
We also conducted the same SED fits for high-$z$ optically-luminous quasars, but no red quasar was found.
This demonstrates the power of Subaru HSC to discover high-$z$ red quasars, which are fainter than the limiting magnitudes of past surveys in the rest-frame ultraviolet, due to dust extinction.

\end{abstract}


\section{Introduction}

\quad High-$z$ quasars (i.e., those with $z > 5.6$) are a useful probe to understand the process of reionization, and the formation and evolution of supermassive black holes (SMBHs) and their host galaxies in the early Universe. 
More than 200 high-$z$ quasars have been discovered to date, by the rest-frame ultraviolet (UV) surveys in all but a few cases (e.g., \citealp{2000AJ....120.1167F, 2001AJ....122.2833F, 2003AJ....125.1649F, 2004AJ....128..515F, 2006AJ....131.1203F, 2006MNRAS.371..769G, 2007AJ....134.2435W, 2009AJ....137.3541W, 2010AJ....139..906W, 2009A&A...505...97M, 2011Natur.474..616M,  2013ApJ...779...24V, 2015MNRAS.453.2259V, 2014AJ....148...14B, 2016ApJS..227...11B, 2018ApJ...856L..25B, 2015MNRAS.454.3952R, 2017MNRAS.468.4702R, 2019MNRAS.487.1874R, 2016ApJ...828...26M, 2018PASJ...70S..35M, 2018ApJS..237....5M, 2019ApJ...872L...2M, 2019ApJ...883..183M, 2016ApJ...833..222J, 2017ApJ...849...91M, 2019ApJ...871..199Y, 2019AJ....157..236Y, 2019ApJ...884...30W}). 
However, the nature of the individual objects is still poorly understood. 
For example, while Atacama Large Millimeter/submillimeter Array (ALMA) observations have revealed the presence of active star formation and abundant dust in the host galaxies (e.g. \citealp{2012ApJ...751L..25V, 2013ApJ...773...44W, 2016ApJ...816...37V, 2018PASJ...70...36I, 2019PASJ...71..111I}), it is unknown to what extent the central quasar radiation is extinguished by the dust. 
This is a critical issue, since the past rest-UV surveys are sensitive only to almost extinction-free quasars;  indeed, a very luminous quasar with $M_{1450} \sim -28$ mag would become as faint as $z_{\rm AB} > 24$ mag with $E(B-V) > 0.3$, and would not be selected even in our deep high-$z$ quasar survey with Subaru Hyper Suprime-Cam (HSC; see below).

This paper focuses on dust-reddened quasars, so-called red quasars, in the high-$z$ universe.
Red quasars have a non-negligible amount of dust extinction, but are not completely obscured.
In this paper, a red quasar means a quasar with color excess of $E(B-V) > 0.1$ \citep{2012ApJ...757...51G}, when the broad-band spectral energy distribution (SED) is fitted with a typical  quasar template and the Small Magellanic Cloud (SMC) dust extinction law over the rest-frame UV/optical wavelengths.
A red quasar is defined to have at least one broad emission line in the spectrum, and thus is different from type 2 quasars.
Type 2 quasars are almost completely obscured in the UV/optical due to our nearly edge-on view of the dust torus, while the sightlines to red quasars may graze the dust torus, or may be obscured by the dust in the host galaxy.
Red quasars have been selected using optical (e.g., \citealp{2003AJ....126.1131R}), optical/mid-infrared (IR) (e.g., \citealp{2015MNRAS.453.3932R, 2017MNRAS.464.3431H}), near-IR/radio (e.g., \citealp{2004ApJ...607...60G, 2007ApJ...667..673G, 2012ApJ...757...51G, 2009ApJ...698.1095U}) and mid-IR techniques (e.g., \citealp{2007AJ....133..186L, 2013ApJS..208...24L, 2018ApJ...861...37G}) at $z < 4$.
Red quasars are currently not known at $z > 5.0$, primarily due to their faintness in the rest-frame UV (observed optical) wavelengths, and to the lack of a large sample of quasars at such high redshifts.

The nature of red quasars, in particular whether they represent a different stage of evolution from normal extinction-free quasars, is still unclear. 
It may be related to the triggering and subsequent transition of quasars - the well-known merger driven scenario suggests that major mergers of gas-rich galaxies first create an obscured starburst phase. 
Obscured active galactic nuclei (AGNs) are triggered at the same time, and may subsequently blow out the interstellar gas of the host galaxies, leading to unobscured quasars, and finally to star formation quenching \citep{1988ApJ...325...74S, 2000MNRAS.311..576K, 2008ApJS..175..356H}.
Red quasars may correspond to the above blowout phase (e.g., \citealp{2012ApJ...757...51G, 2009ApJ...698.1095U}). 
Other mechanisms have also been suggested to trigger AGNs, such as disc instability \citep{1998MNRAS.295..319M, 2000MNRAS.319..168C, 2006MNRAS.370..645B, 2008MNRAS.388..587L, 2011MNRAS.410...53F}, minor mergers \citep{1995ApJ...448...41H, 1998ApJ...496...93D, 2008MNRAS.388..587L} or gas feeding from ordinary stellar processes \citep{2007ApJ...665.1038C}.  
Red quasars may or may not represent a distinct stage of these processes.  
Constructing a large sample of red quasars at various redshifts and environments may help to understand the physical conditions required to produce this population.

This is the ninth paper from the Subaru High-$z$ Exploration of Low-Luminosity Quasars (SHELLQs) project \citep{2016ApJ...828...26M}, based on the Subaru HSC survey data. 
We aim to identify red quasars from the HSC high-$z$ quasars, with the aid of IR photometry from Wide-field Infrared Survey Explore ($\it WISE$; \citealp{2010AJ....140.1868W}).
We compiled the optical, near-IR, and $\it WISE$ ($W1, W2$) photometry of the quasars, and performed SED fitting to look for evidence of dust reddening.
We note that even with the two $\it WISE$ bands, we can trace the rest-frame spectral coverage only up to $\sim$6500 \AA, and cannot access near-IR portion of a spectrum.
As we already mentioned, given that the quasars were selected in the rest-frame UV, we are sensitive only to modest amounts of extinction.
Nonetheless, the unprecedented depth of the HSC survey has a potential of finding red quasars that were missed in the previous surveys.

This paper is structured as follows.
We introduce the data and sample in Section 2.
In Section 3, we perform image decomposition of blended $\it WISE$ sources, and then broad-band SED fitting with a quasar template.
Our results are discussed in Section 4, and summarized in Section 5.
This paper adopts the cosmological parameters $H_{0}$ = 70 km s$^{-1}$ Mpc$^{-1}$, $\Omega_{\rm M}$ = 0.3, and $\Omega_{\rm \Lambda}$ = 0.7.
All magnitudes are presented in the AB system \citep{1983ApJ...266..713O}.

\section{Data and Sample}

\quad  The SHELLQs project is based on the sensitive multi-band optical imaging data produced by the HSC Subaru Strategic Program (SSP) survey\footnote{\url{https://hsc.mtk.nao.ac.jp/ssp/}} \citep{2018PASJ...70S...4A}. 
The HSC is a wide-field optical imaging camera installed on the 8.2 m Subaru Telescope on the summit of Maunakea, and covers a 1.5 degree diameter field of view \citep{2018PASJ...70S...1M, 2018PASJ...70S...2K, 2018PASJ...70...66K, 2018PASJ...70S...3F}.  
The HSC-SSP is an imaging survey with five broad bands ($g, r, i, z,$ and $y$) plus several narrow bands, and has three layers with different combinations of area and depth (Wide, Deep, and UltraDeep). 
The survey started in March 2014, and will cover 1400 $\rm deg^2$ in the Wide layer \citep{2018PASJ...70S...4A} when completed. 
The 5$\sigma$ depths for a point source are ($g, r, i, z, y$) = (26.5, 26.1, 25.9, 25.1, 24.4) mag in the Wide layer.
The HSC data are processed with a pipeline that measures the properties of all detected objects \citep{2018PASJ...70S...5B}.

The SHELLQs project has spectroscopically confirmed 93 low-luminosity quasars at $z > 5.6$ selected from the HSC data \citep{2016ApJ...828...26M, 2018PASJ...70S..35M, 2018ApJS..237....5M, 2019ApJ...872L...2M, 2019ApJ...883..183M}. 
This includes the first low-luminosity quasar at $z > 7$ \citep{2019ApJ...872L...2M}.
The SHELLQs quasars were selected by a Bayesian probabilistic algorithm, and like other high-$z$ quasar surveys, do not include quasars whose colors are close to those of Galactic stars.
More specifically, the quasars were initially selected as point sources satisfying:
$$ z_{\rm PSF} < 24.5\ \&\ \sigma_z < 0.155\  \&\ i_{\rm PSF}-z_{\rm PSF} > 1.5 $$
$$ \&\ z_{\rm PSF}-z_{\rm CModel} < 0.15 \eqno(1) $$
or
$$ y_{\rm PSF} < 25.0\ \&\ \sigma_y < 0.217\  \&\ z_{\rm PSF}-y_{\rm PSF} > 0.8 $$
$$ \&\ y_{\rm PSF}-y_{\rm CModel} < 0.15. \eqno(2) $$
Here $m_{\rm PSF}$ is the point spread function (PSF) magnitude measured by fitting the PSF model to a given source, while $m_{\rm CModel}$ is the CModel magnitude measured by fitting PSF-convolved, weighted combination of exponential and de Vaucouleurs models to the source \citep{2018PASJ...70S...5B}. 
Equations (1) and (2) represent the criteria of $i$-band dropout and $z$-band dropout candidates, respectively, where $\sigma_m$ is the error of the PSF magnitude.
The Bayesian algorithm uses sky surface density and spectral models of high-z quasars and contaminating brown dwarfs, and calculates probability ($P_{\rm Q}$) for each source being a high-z quasar. We are carrying out spectroscopic follow-up of those candidates meeting $P_{\rm Q} > 0.1$. 
We are also carrying out multi-wavelength follow-up observations of the discovered quasars; \citet{2018PASJ...70...36I, 2019PASJ...71..111I} report the results from ALMA observations, and \citet{2019ApJ...880...77O} present initial results from deep near-IR spectroscopy of several objects.

This paper uses $\it WISE$ data to select red quasars.
The $\it WISE$ bands provide rest-frame spectral coverage at $> 3500$ \AA, which is inaccessible with common $JHK$ bands, for $z > 6$ sources.
The $\it WISE$ performed an all-sky imaging survey in the 3.4 ($W1$), 4.6 ($W2$), 12 ($W3$), and 22 ($W4$) $\mu$m bands, with the angular resolution of 6.1, 6.4, 6.5, and 12.0 arcsec, respectively. 
The 5$\sigma$ depths for a point source are ($W1, W2, W3, W4$) = (19.6, 19.3, 16.8, 14.7) AB mag\footnote{\url{http://wise2.ipac.caltech.edu/docs/release/allwise/expsup/sec2_3a.html}}.
In this study, we use the AllWISE Source Catalog\footnote{\url{http://wise2.ipac.caltech.edu/docs/release/allwise/expsup/sec1_1.html}}. 
This catalog is superior to the WISE All-Sky Release Catalog\footnote{\url{http://wise2.ipac.caltech.edu/docs/release/allsky}} in the $W1$ and $W2$ bands, with improved photometric/astrometric accuracy. 
The magnitudes were converted from the Vega to AB system by adding 2.699 ($W1$), 3.339 ($W2$), 5.174 ($W3$), and 6.620 ($W4$) to the catalog values\footnote{\url{http://wise2.ipac.caltech.edu/docs/release/allsky/expsup/sec4_4h.html}}.

\begin{figure}[t]
 \begin{center}

       \includegraphics[clip, width=8.cm]{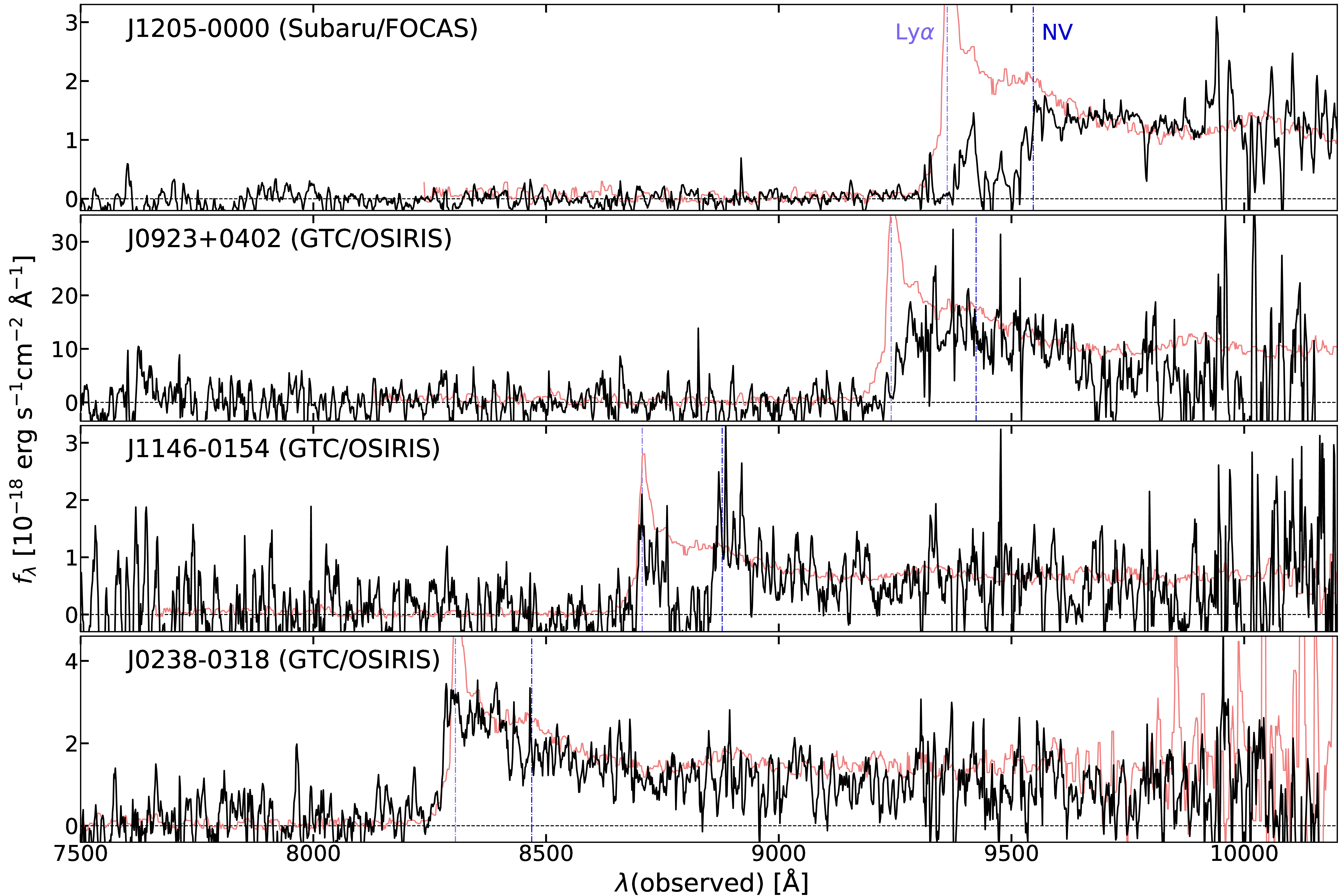} 
    
 \end{center}
\caption{Optical spectra (black line) of the red quasar candidates observed by Subaru/FOCAS or GTC/OSIRIS. 
The spectra were smoothed using inverse-variance weighted means over 5 pixels.
The red line represents the stack spectrum of the SHELLQs quasars with unambiguous broad lines. 
The vertical lines represent the central wavelength of broad emission lines (Ly$\alpha$ (purple) and NV (blue)).
J1205$-$0000 and J1146$-$0154 have BAL features at bluewards of NV lines.}
\label{fig:optical_spectrum}
\end{figure}

\begin{figure}[t]
 \begin{center}
  \begin{tabular}{l}

  $\triangleright$ J1205$-$0000\\
  	
       \includegraphics[clip, width=7.8cm]{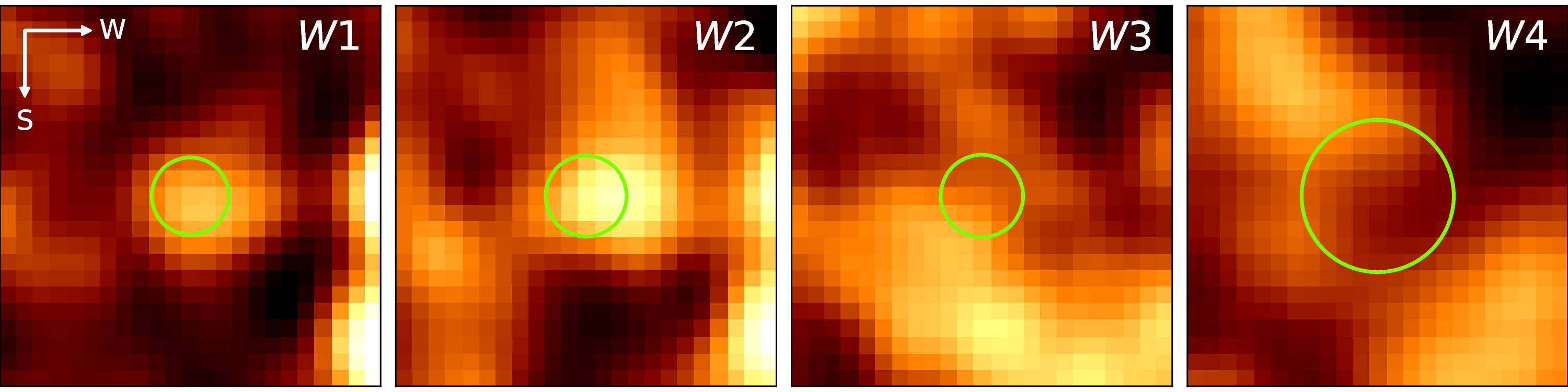} \\

  $\triangleright$ J0923$+$0402\\

       \includegraphics[clip, width=7.8cm]{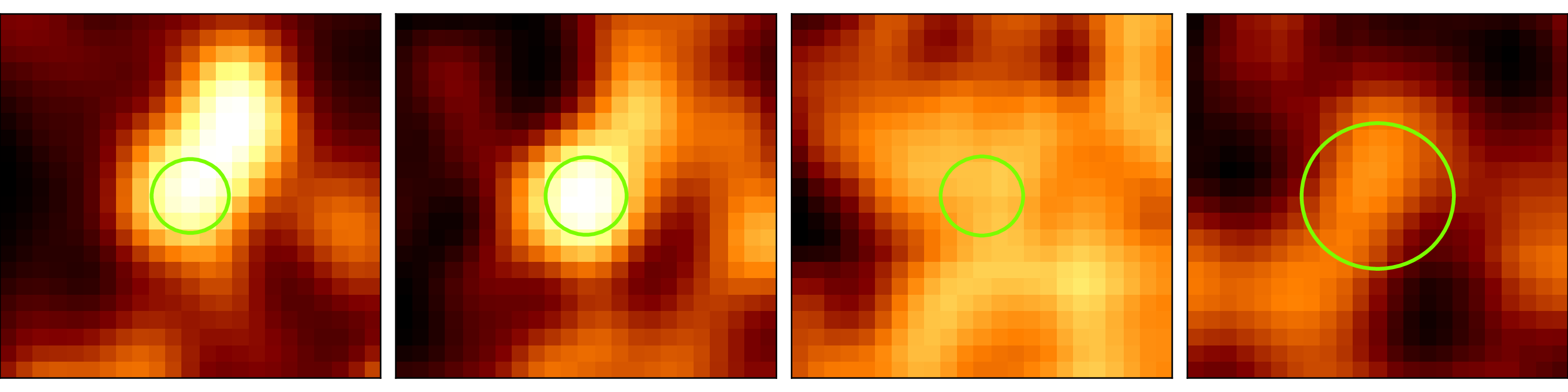} \\
    
  $\triangleright$ J1146$-$0154\\

       \includegraphics[clip, width=7.8cm]{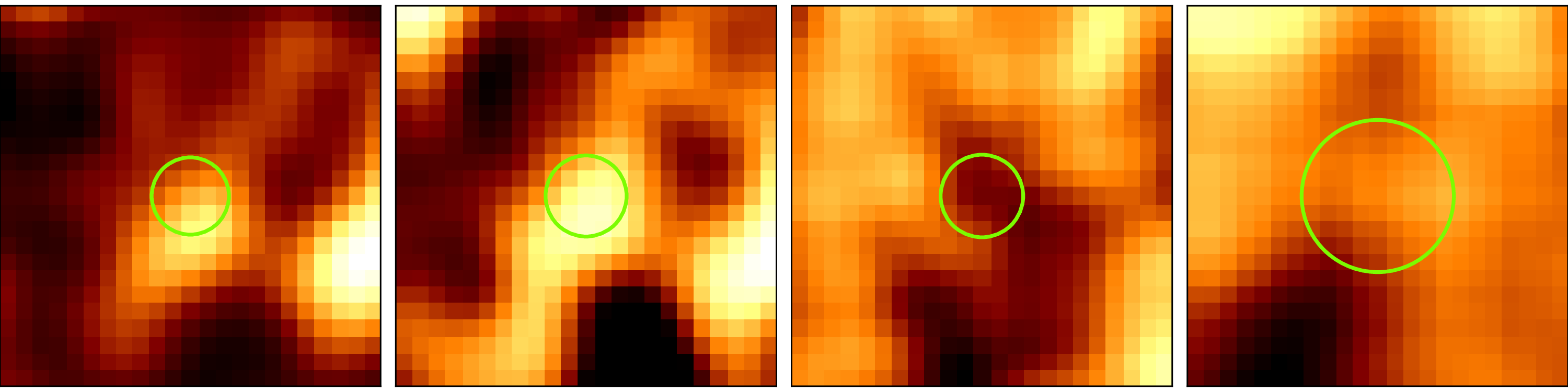} \\
    
  $\triangleright$ J0238$-$0318\\

       \includegraphics[clip, width=7.8cm]{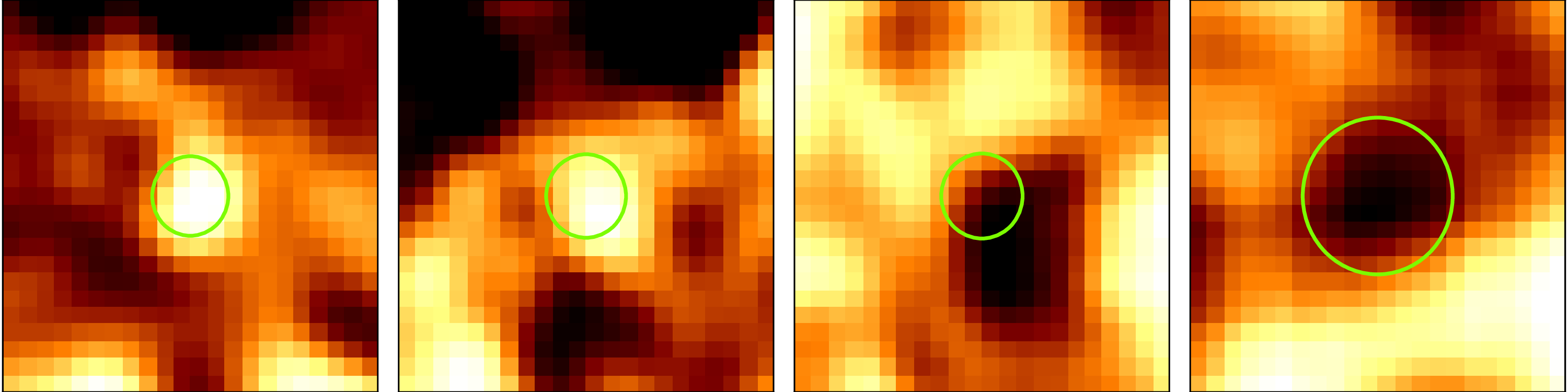} \\
    
    \end{tabular}
 \end{center}
\caption{The $\it WISE$ images of the red quasar candidates ($W1$, $W2$, $W3$, and $W4$ band from left to right). 
The image size is 30 arcsec on a side.
The green circles indicate the angular resolution; 6.1, 6.4, 6.5, and 12 arcsec in $W1$, $W2$, $W3$, and $W4$, respectively.}
\label{fig:WISE_image}
\end{figure}

We initially selected red quasar candidates by matching the SHELLQs sample to the $\it WISE$ sources, with a matching radius of 3.0 arcsec. 
Note that the $W2$ band is shallower by 0.3 mag than the $W1$, which is a good match to the color ($W1 - W2 = 0.3$) of a typical quasar spectrum \citep{2016A&A...585A..87S} with $E(B-V) = 0.1$.
We found that four SHELLQs quasars - J120505.09$-$000027.9 (J1205$-$0000), J092347.12$+$040254.5 (J0923$+$0402), J114632.66$-$015438.2 (J1146$-$0154), and J023858.09$-$031845.4 (J0238$-$0318) - have $\it WISE$ counterparts in the $W1$ and $W2$ bands (S/N $=$ 2--15), and regarded these four quasars as red quasar candidates for further analysis. 
Figure \ref{fig:optical_spectrum} displays their optical spectra, obtained with the Subaru/Faint Object Camera and Spectrograph (FOCAS; \citealp{2002PASJ...54..819K}) or the Gran Telescopio Canarias (GTC)/Optical System for Imaging and low-Intermediate-Resolution Integrated Spectroscopy (OSIRIS; \citealp{2000SPIE.4008..623C}) \citep{2016ApJ...828...26M, 2018ApJS..237....5M, 2019ApJ...883..183M}. 
We note that two of the four candidates, J1205$-$0000 and J1146$-$0154, have broad absorption line (BAL) features.
In Figure \ref{fig:WISE_image}, we show the $\it WISE$ images of the candidates. 
Table \ref{table:candidates} presents their photometric and spectroscopic properties. 
None of the candidates are detected in the $W3$ or $W4$ band.
Note that, since the surface density of $\it WISE$ sources in the HSC-Wide layer is $\sim$2.6 arcmin$^{-2}$, a single aperture with 3.0 arcsec radius would contain a $\it WISE$ source by chance with $\sim1.9\%$ probability. 
Thus the 93 quasars are expected to have 1.7 matched $\it WISE$ objects by pure chance coincidence.

\renewcommand{\thefootnote}{\fnsymbol{footnote}}
\begin{table*}[t]
 \caption{Red quasar candidates}
 \label{table:candidates}
 \centering
\scalebox{0.85}{
  \begin{tabular}{lcccccccccc}
  \hline\hline
  
  Object & $i_{\rm AB}$(mag) & $z_{\rm AB}$(mag) & $y_{\rm AB}$(mag) & $W1_{\rm AB}$(mag) & $W2_{\rm AB}$(mag) & Redshift & $M_{1450}$(mag) & ref\\
  \hline
  HSC J120505.09$-$000027.9	& $> 26.72$ 		& $> 25.94$ & $22.60 \pm 0.03$ & $19.98 \pm 0.15$ & $19.65 \pm 0.23$ & \ 6.70\footnote[2]  & $-24.56 \pm 0.04$ & (1)\\
  HSC J092347.12$+$040254.5& $26.27 \pm 0.20$	& $22.64 \pm 0.02$ & $20.21 \pm 0.01$ & $19.06 \pm 0.07$ & $19.20 \pm 0.16$ & 6.60 & $-26.18 \pm 0.14$ & (2)\\
  HSC J114632.66$-$015438.2 	& $26.57 \pm 0.31$	& $23.63 \pm 0.06$ & $23.78 \pm 0.15$ & $20.04 \pm 0.16$ & $20.16 \pm 0.38$ & 6.16 & $-23.43 \pm 0.07$ & (2)\\
  HSC J023858.09$-$031845.4 & $24.17 \pm 0.08$	& $22.64 \pm 0.04$ & $22.53 \pm 0.08$ & $20.57 \pm 0.21$ & $20.71 \pm 0.50$ & 5.83 & $-23.94 \pm 0.03$ & (3)\\  
  \hline
\end{tabular}
}\\
\begin{tabnote}
\raggedright
\textbf{Notes. }
--- The coordinates are at J2000.0. The HSC coordinates are tied to the astrometric catalog from Gaia \citep{2019PASJ...71..114A}. The HSC magnitudes (psfMag) were taken from the S18A internal data release. The magnitude lower limits are placed at $5\sigma$ significance. The $\it WISE$ magnitudes were taken from the AllWISE Source Catalog. References : (1) \citet{2018PASJ...70S..35M}, (2) \citet{2018ApJS..237....5M}, (3) \citet{2019ApJ...883..183M} \\
\dag \ The redshift of J1205$-$0000 was measured from the near-IR spectrum presented in \citet{2019ApJ...880...77O} ($z = 6.699^{+0.007}_{-0.001}$).
\end{tabnote}
\end{table*}

\section{Analysis and Results}

\begin{figure}[t]
 \begin{center}
  \begin{tabular}{c}

    \begin{minipage}{0.23\hsize}
      \begin{center}
       \hspace{0.08cm} J1205$-$0000
       \includegraphics[clip, width=1.95cm]{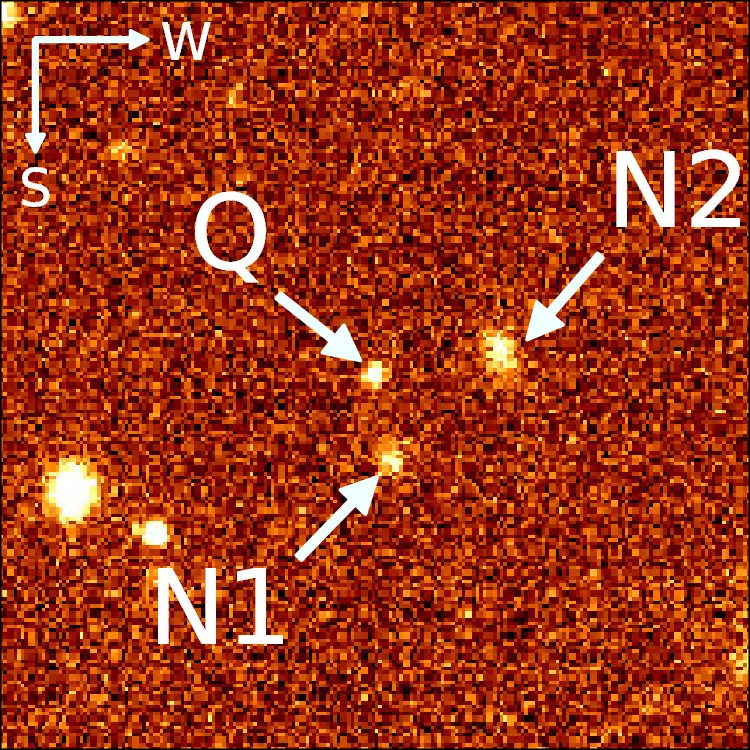} 
      \end{center}
    \end{minipage}
    
    \begin{minipage}{0.23\hsize}
      \begin{center}
       \hspace{0.08cm} J0923$+$0402
       \includegraphics[clip, width=1.95cm]{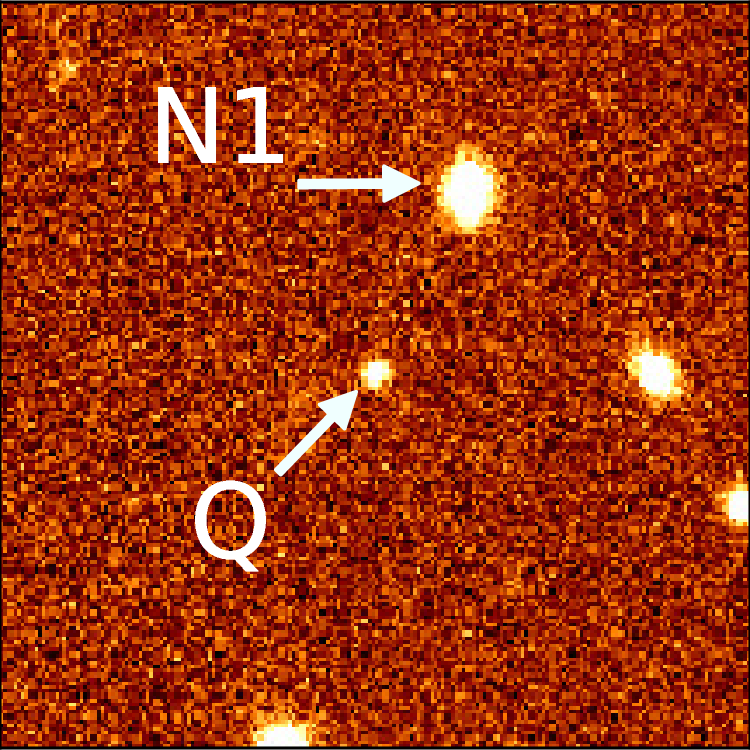} 
      \end{center}
    \end{minipage}
 
    \begin{minipage}{0.23\hsize}
      \begin{center}
       \hspace{0.08cm} J1146$-$0154
       \includegraphics[clip, width=1.95cm]{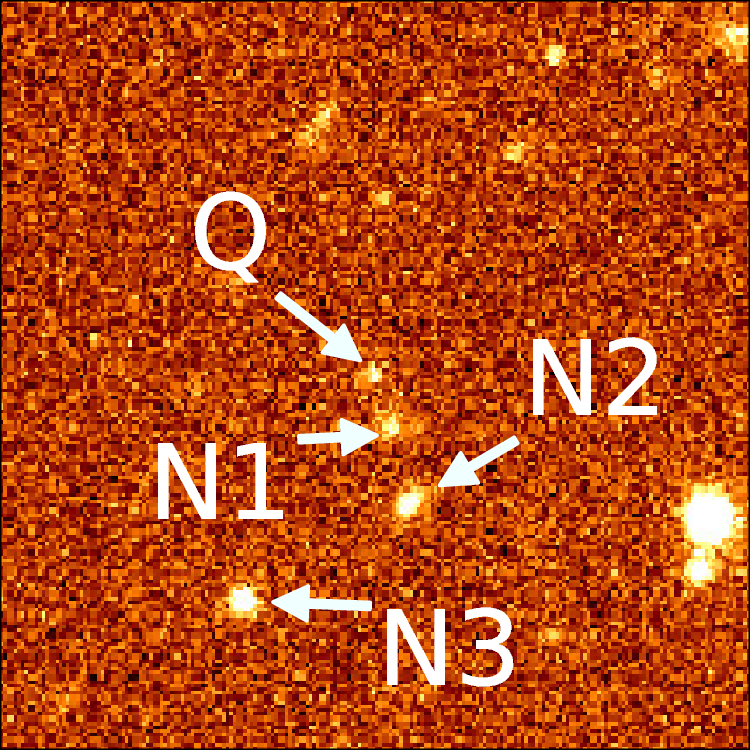} 
      \end{center}
    \end{minipage}
    
    \begin{minipage}{0.23\hsize}
      \begin{center}
       \hspace{0.08cm} J0238$-$0318
       \includegraphics[clip, width=1.95cm]{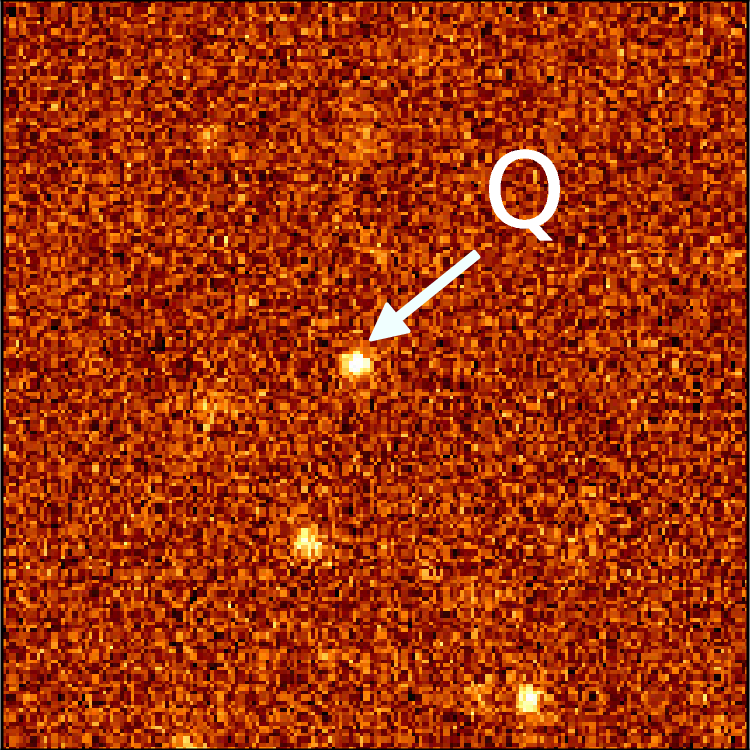} 
      \end{center}
    \end{minipage}\\
    
 \end{tabular}
 \end{center}
\caption{The HSC images of the red quasar candidates ($y$-band for J1205$-$0000, and $z$-band for the other three objects). 
The image size is 30 arcsec on a side. 
"Q" and "N" mark the quasar and nearby objects identified on the HSC images which could contribute to the AllWISE catalog flux.
These are the HSC-Wide images with the 5$\sigma$ limiting magnitudes of $y \sim 24.4$ mag (J1205$-$0000) and $z \sim25.1$ mag (the remaining panels).}
\label{fig:HSC_image}
\end{figure}

\begin{figure}[t]
 \begin{center}
  \begin{tabular}{c}

    \begin{minipage}{0.23\hsize}
      \begin{center}
       \hspace{0.08cm} J1205$-$0000
       \includegraphics[clip, width=1.95cm]{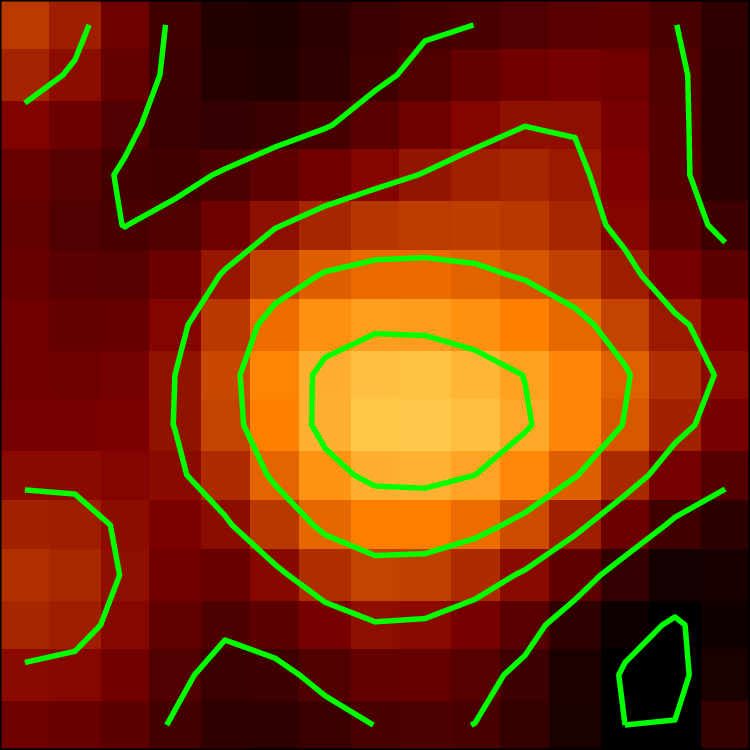} 
      \end{center}
    \end{minipage}
    
    \begin{minipage}{0.23\hsize}
      \begin{center}
       \hspace{0.08cm} J0923$+$0402
       \includegraphics[clip, width=1.95cm]{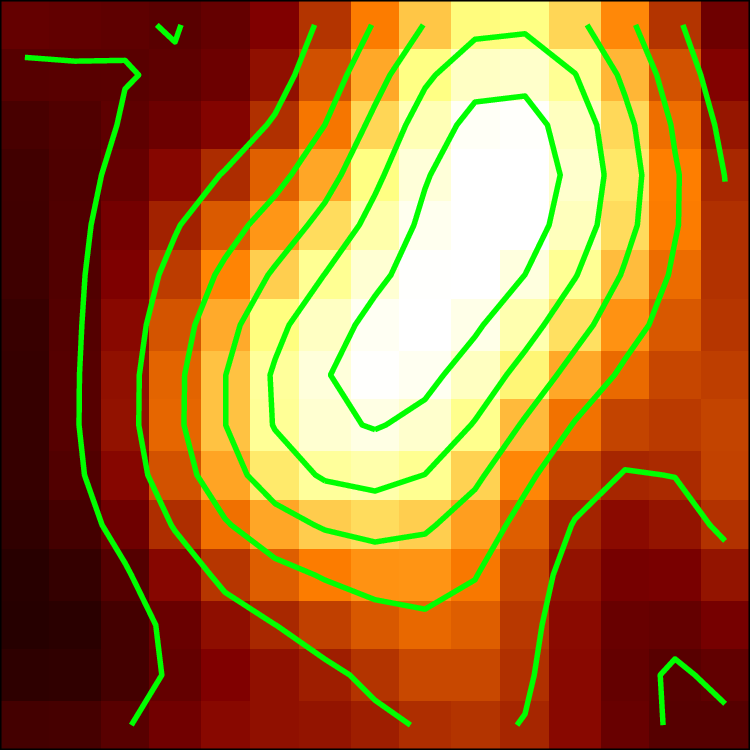} 
      \end{center}
    \end{minipage}
 
    \begin{minipage}{0.23\hsize}
      \begin{center}
       \hspace{0.08cm} J1146$-$0154
       \includegraphics[clip, width=1.95cm]{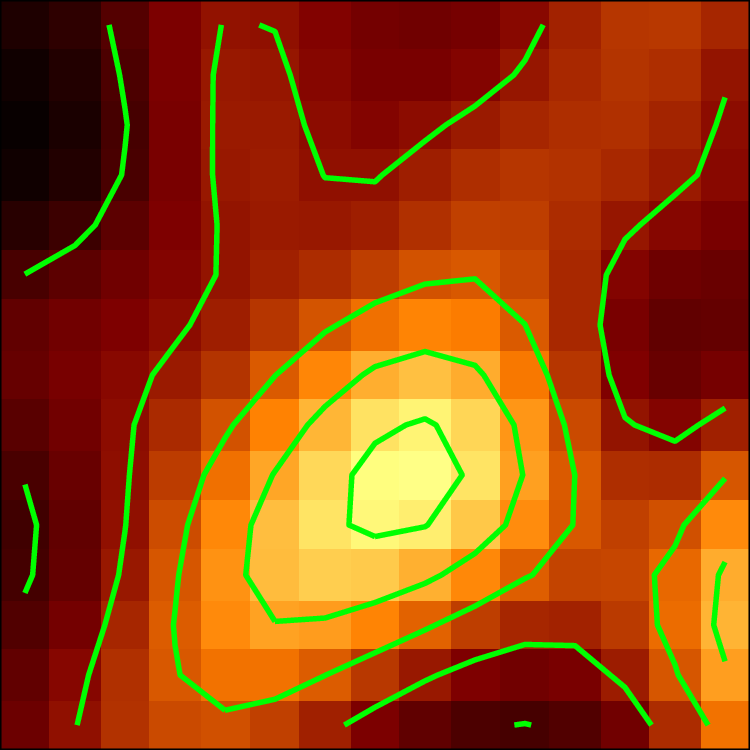} 
      \end{center}
    \end{minipage}
    
    \begin{minipage}{0.23\hsize}
      \begin{center}
       \hspace{0.08cm} J0238$-$0318
       \includegraphics[clip, width=1.95cm]{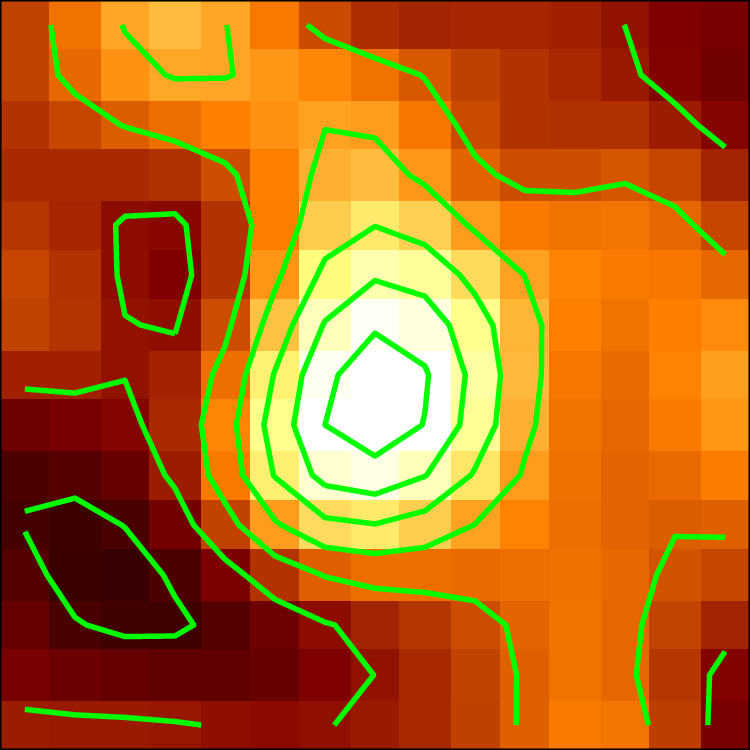} 
      \end{center}
    \end{minipage}\\

    \begin{minipage}{0.23\hsize}
      \begin{center}
       \includegraphics[clip, width=1.95cm]{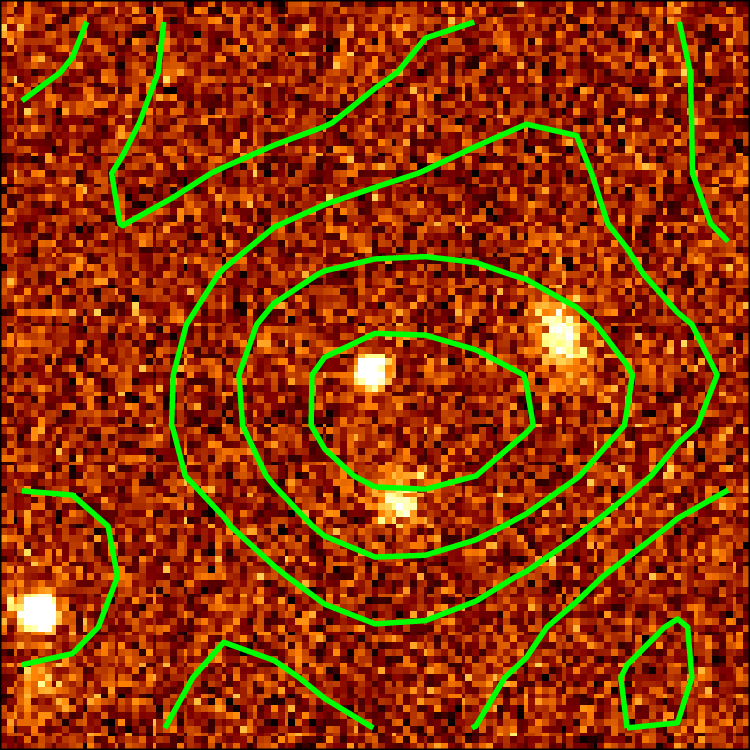} 
      \end{center}
    \end{minipage}

    \begin{minipage}{0.23\hsize}
      \begin{center}
       \includegraphics[clip, width=1.95cm]{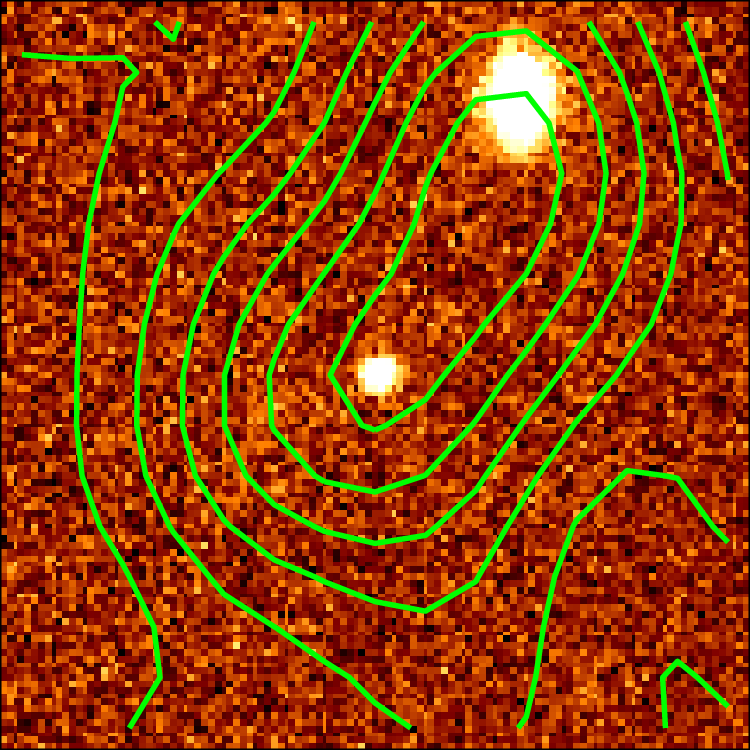} 
      \end{center}
    \end{minipage}

    \begin{minipage}{0.23\hsize}
      \begin{center}
       \includegraphics[clip, width=1.95cm]{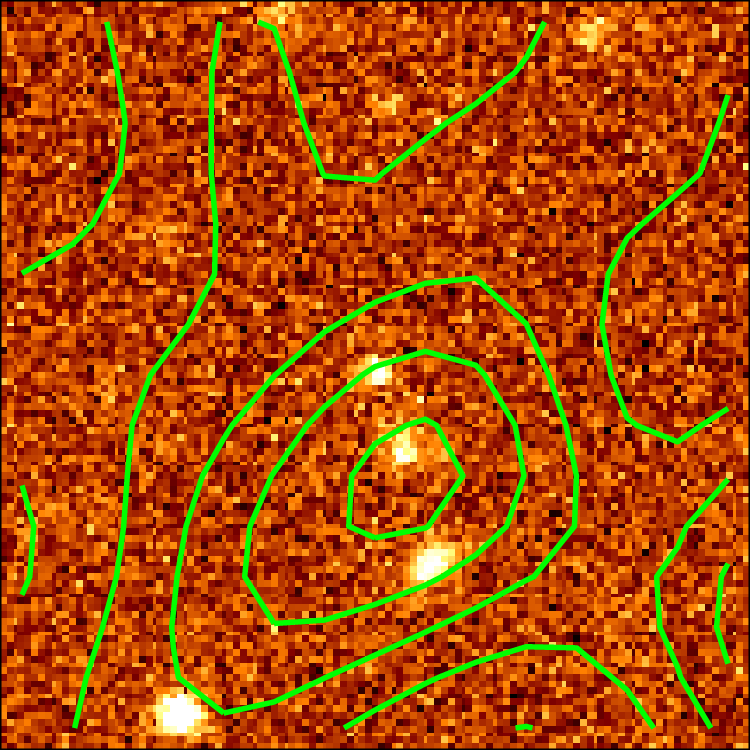} 
      \end{center}
    \end{minipage}

    \begin{minipage}{0.23\hsize}
      \begin{center}
       \includegraphics[clip, width=1.95cm]{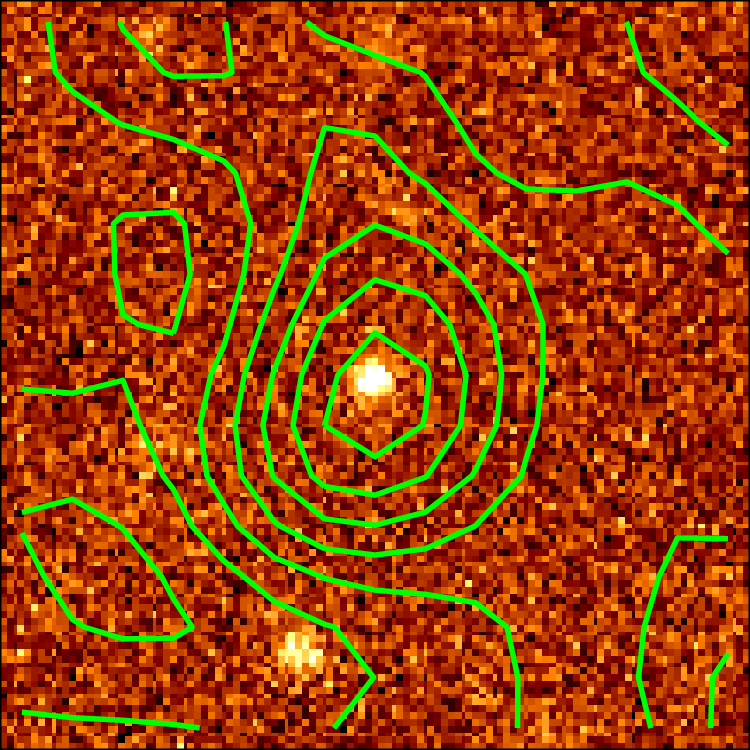} 
      \end{center}
    \end{minipage}\\
    
    \begin{minipage}{0.23\hsize}
      \begin{center}
       \includegraphics[clip, width=1.95cm]{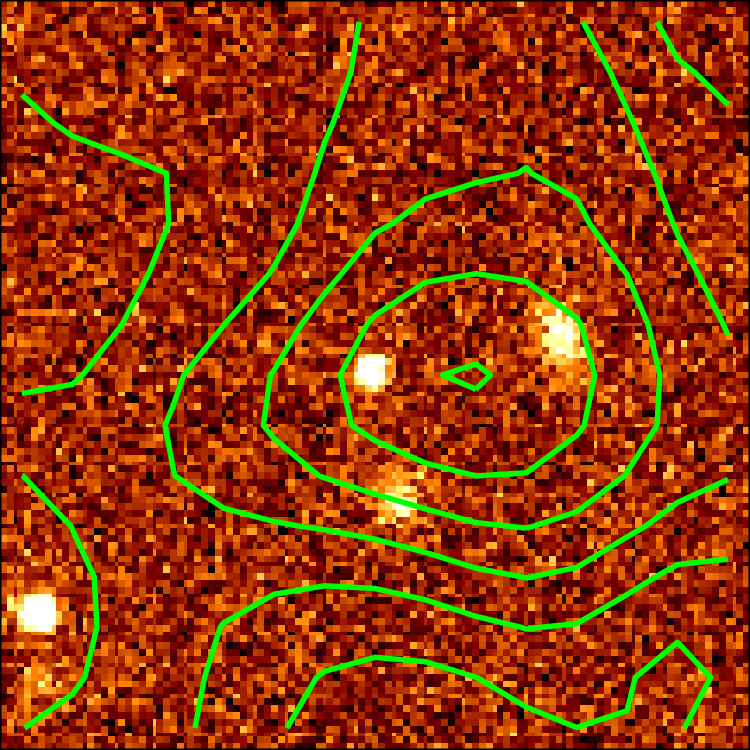} 
      \end{center}
    \end{minipage}

    \begin{minipage}{0.23\hsize}
      \begin{center}
       \includegraphics[clip, width=1.95cm]{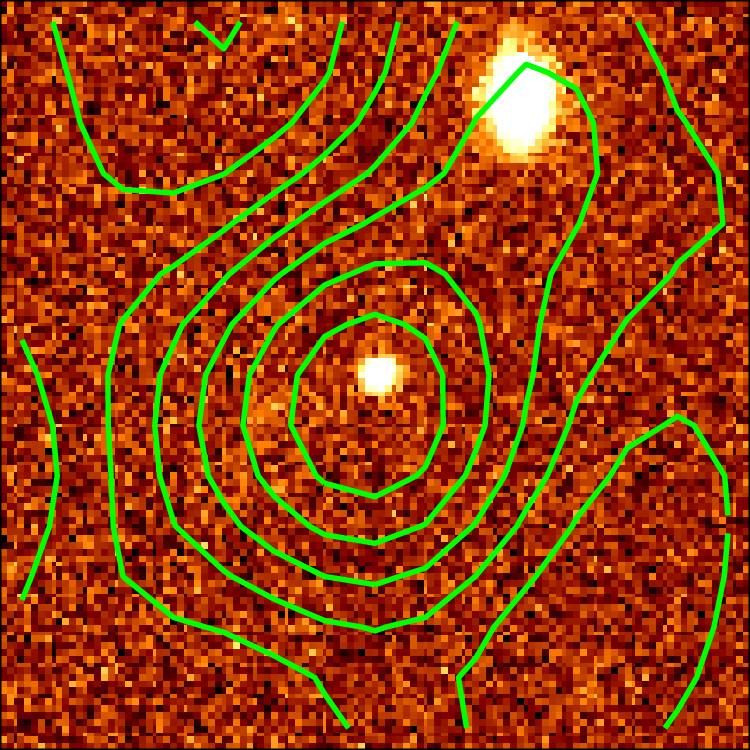} 
      \end{center}
    \end{minipage}

    \begin{minipage}{0.23\hsize}
      \begin{center}
       \includegraphics[clip, width=1.95cm]{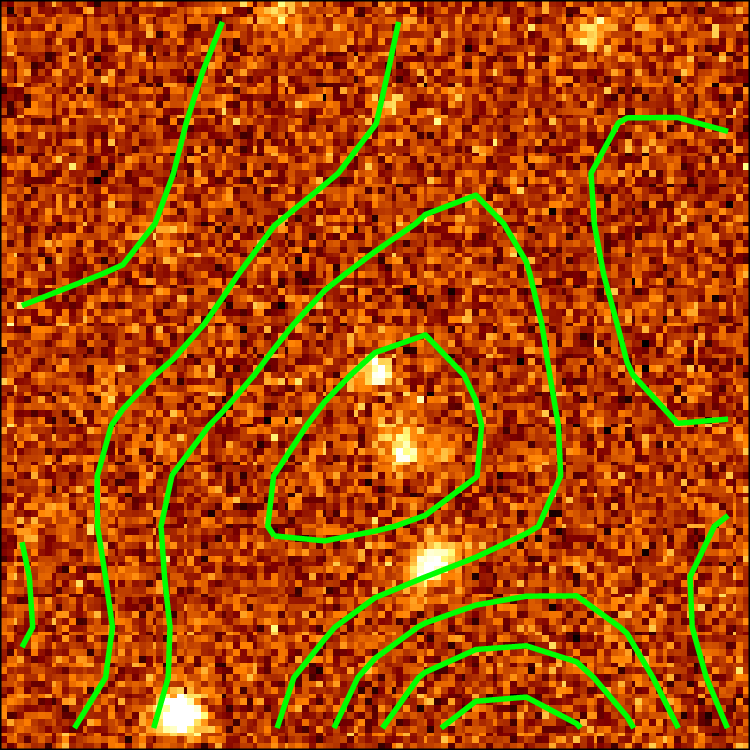} 
      \end{center}
    \end{minipage}

    \begin{minipage}{0.23\hsize}
      \begin{center}
       \includegraphics[clip, width=1.95cm]{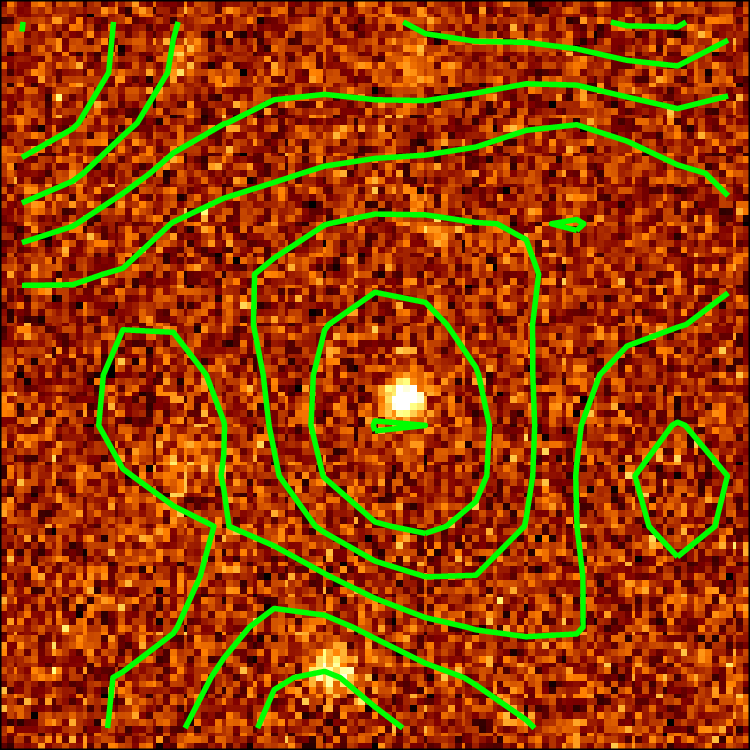} 
      \end{center}
    \end{minipage}\\

 \end{tabular}
 \end{center}
\caption{$W1$ images (top) and HSC images with $W1/W2$ flux contours overlaid with green lines (middle/bottom) around the four red quasar candidates.
The image size is 20 arcsec on a side.
Note that the astrometric uncertainty of the $\it WISE$ data is much smaller than the $\it WISE$ PSF sizes.}
\label{fig:contour_image}
\end{figure}

\quad As shown in Figures \ref{fig:HSC_image} and \ref{fig:contour_image}, we found neighbors around three quasars (J1205$-$0000, J0923$+$0402, and J1146$-$0154) in the HSC images, which may contribute to the AllWISE catalog fluxes.
Therefore, we modeled and decomposed the $\it WISE$ images as a superposition of PSFs placed at the positions of the HSC detections.
Here we assume that the neighbors are unresolved in $\it WISE$ images, which indeed gives good fits to the observed $\it WISE$ images, as we will see below.

\begin{figure*}[t]

\raggedright
$\triangleright$ J1205$-$0000 \\

\begin{tabular}{ccc}
	\begin{minipage}{.48\textwidth}
    		\begin{center}
		\includegraphics[clip, width=8.5cm]{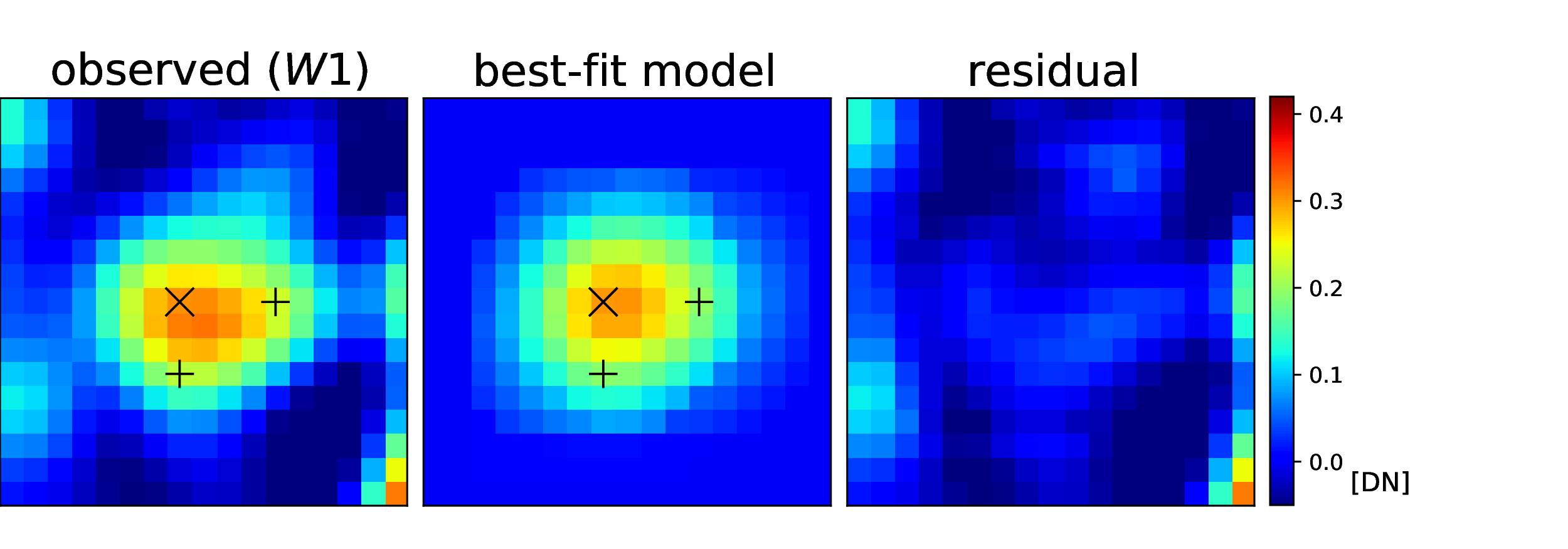} 
		\end{center}
	\end{minipage}
	
	\begin{minipage}{.48\textwidth}
    		\begin{center}
		\includegraphics[clip, width=8.5cm]{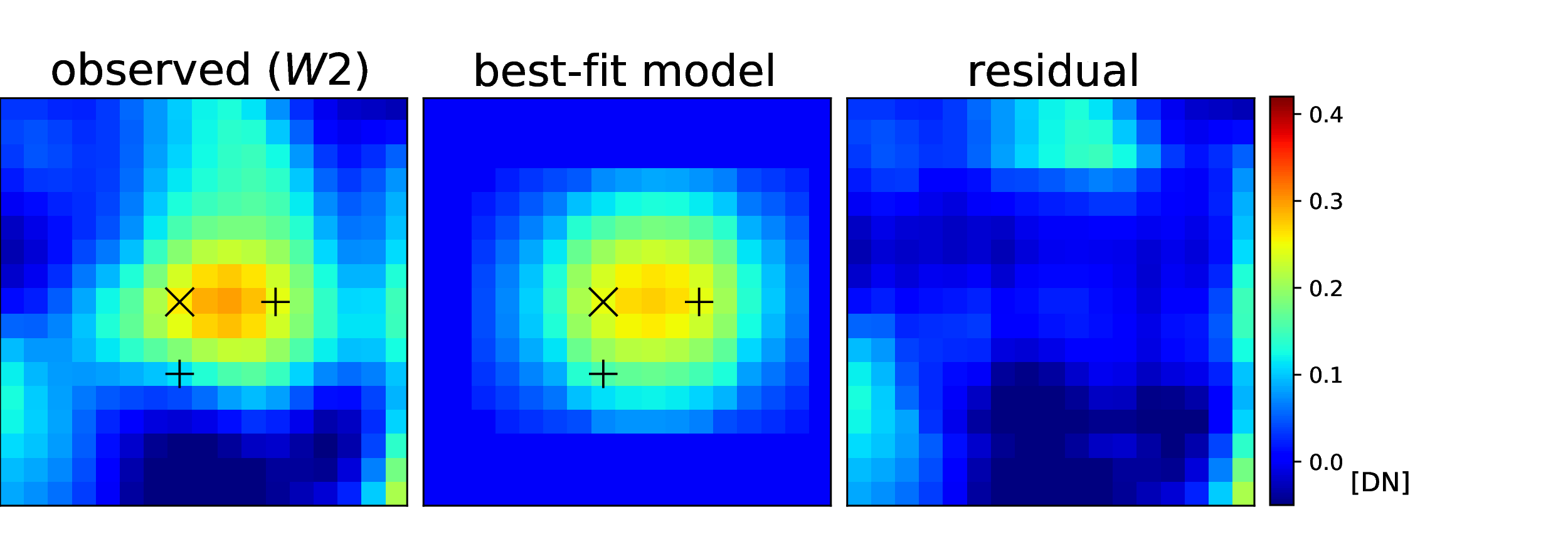} 
		\end{center}
	\end{minipage}
\end{tabular}

\raggedright
$\triangleright$ J0923$+$0402 \\

\begin{tabular}{ccc}
	\begin{minipage}{.48\textwidth}
    		\begin{center}
		\includegraphics[clip, width=8.5cm]{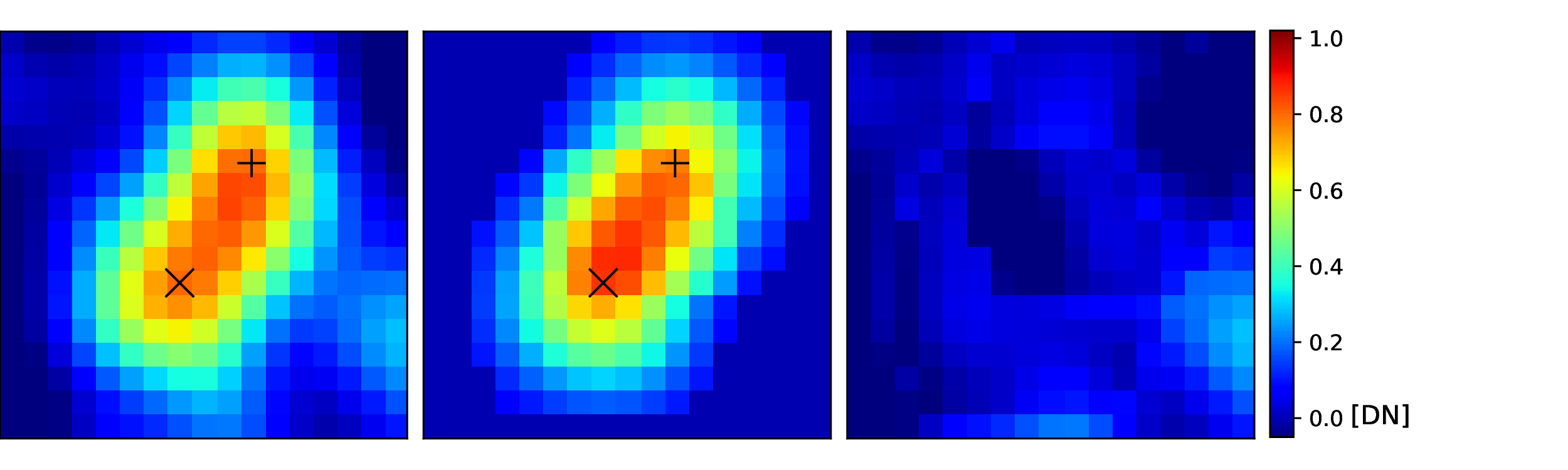} 
		\end{center}
	\end{minipage}
	
	\begin{minipage}{.48\textwidth}
    		\begin{center}
		\includegraphics[clip, width=8.5cm]{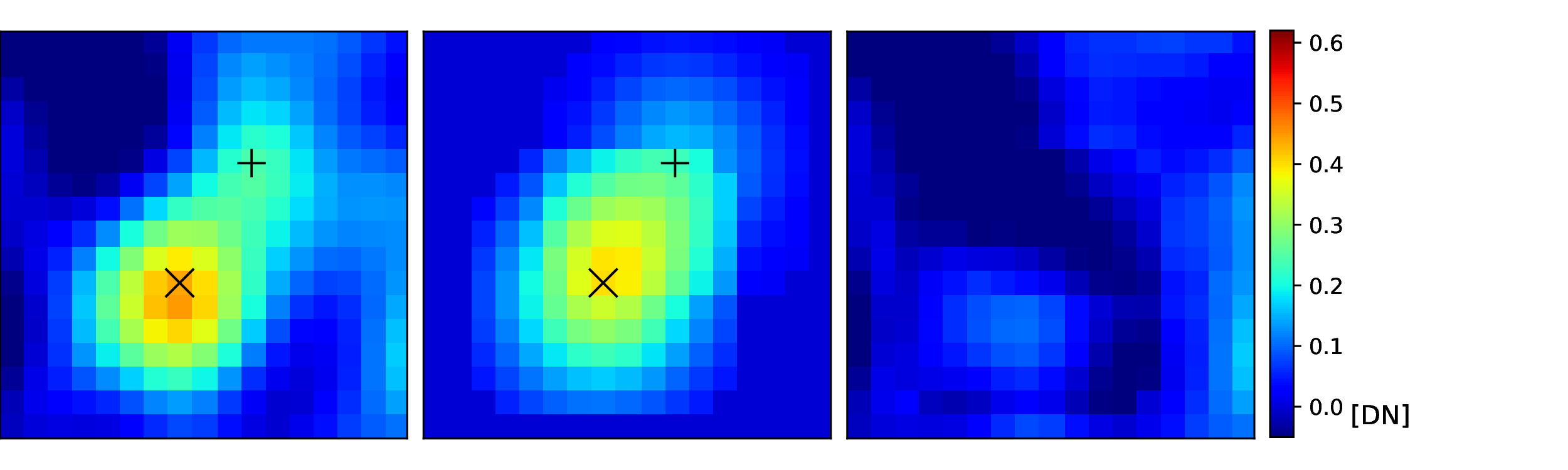} 
		\end{center}
	\end{minipage}
\end{tabular}

\raggedright
$\triangleright$ J1146$-$0154 \\

\begin{tabular}{ccc}
	\begin{minipage}{.48\textwidth}
    		\begin{center}
		\includegraphics[clip, width=8.5cm]{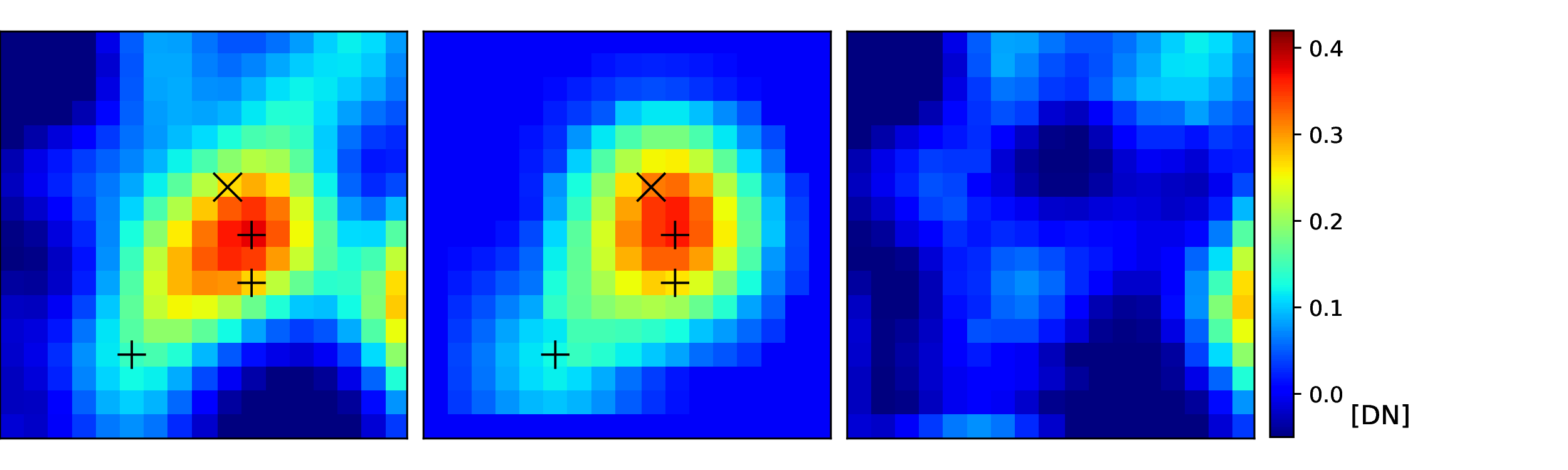} 
		\end{center}
	\end{minipage}
	
	\begin{minipage}{.48\textwidth}
    		\begin{center}
		\includegraphics[clip, width=8.5cm]{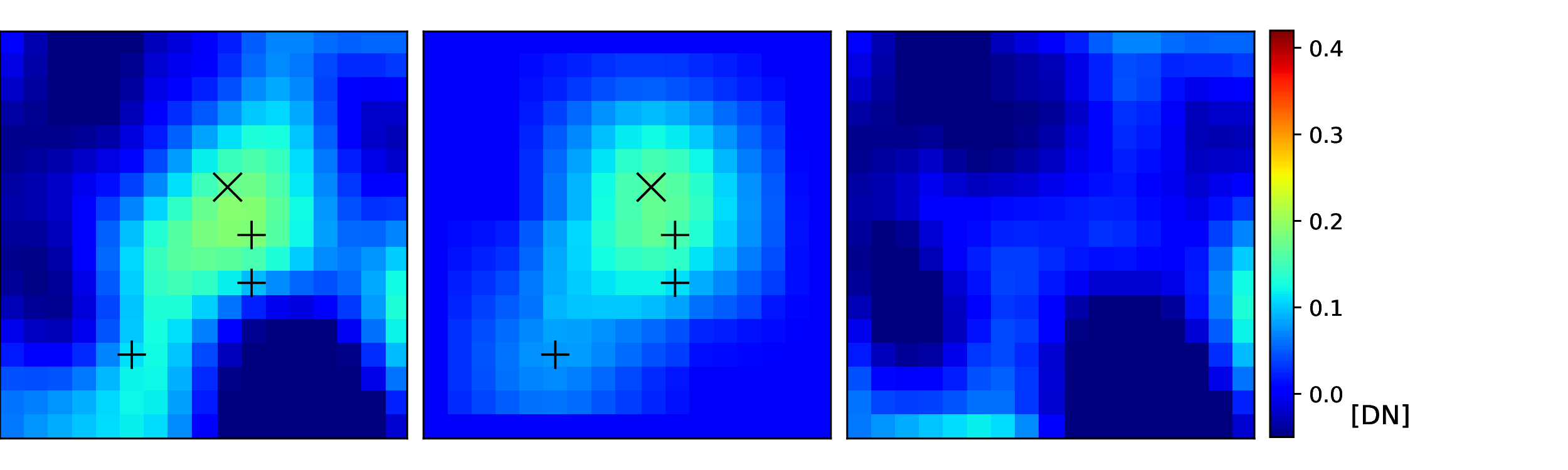} 
		\end{center}
	\end{minipage}
\end{tabular}

\caption{$\it WISE$ image decomposition.
The left part shows the observed image (left), the best-fit model (middle), and the residual (right) in the $W1$ band.
The right part shows those in the $W2$ band.
The cross and plus symbols represent the positions of the quasar and the neighbors, respectively. 
The image size is 23 arcsec on a side.
J0238$-$0318 do not have neighbors contributing to the quasar $\it WISE$ flux, and thus is not included here.}
\label{fig:modeling_results}	
\end{figure*}

To model the $\it WISE$ PSFs, we mean-stacked $\sim$10 bright and isolated point sources around each quasar found in the $\it WISE$ images, after normalizing the individual source profiles at the peak.
We then superposed the scaled PSFs placed at the HSC positions of the quasar and neighbors, and performed $\chi^2$ fitting to the observed $\it WISE$ ($W1, W2$) images, looking for the best-fit scaling parameters for the individual PSFs. 
We modeled the $W1$ and $W2$ images independently.
The decomposed $\it WISE$ quasar flux is then obtained by integrating the corresponding best-fit PSF.
Figure \ref{fig:modeling_results} presents the modeled $\it WISE$ images of the three candidates.
The decomposed flux errors were computed by combining in quadrature the AllWISE catalog flux errors (reflecting the background and calibration uncertainties) and the errors coming from the decomposition process (which are of the order of $\sim$10\% of the flux), the latter being estimated from the $\chi^2$ distributions.
Since J0238$-$0318 has no neighbors that seem to contribute to its $\it WISE$ flux (see the bottom-right panel of Figure \ref{fig:contour_image}), we simply adopt its AllWISE catalog fluxes and errors.

The color excess $E(B-V)$ of the four quasars were estimated via broad-band SED fitting. 
We used the photometric data in the optical (HSC), near-IR, and mid-IR ($\it WISE$) bands. 
We do not use the $W3$- and $W4$-band information, since the relatively poor sensitivity does not allow us to give meaningful flux upper limits in these bands.
The near-IR magnitudes were obtained from the VISTA Kilo-Degree Infrared Galaxy Survey (VIKING; \citealt{2007Msngr.127...28A}) and the UK Infrared Telescope Infrared Deep Sky Survey (UKIDSS; \citealt{2007MNRAS.379.1599L}) using a matching radius of 1 arcsec (see Table \ref{table:mag_info}). 
The near-IR surveys have the angular resolution of $\sim$1 arcsec, and thus the quasar photometry is not affected by nearby objects.
In principle, dust extinction makes a quasar spectrum flatter, while such flattening would also happen when there is a significant contribution from the host galaxy. 
We used two models for the SED fitting. 
The first model is a typical quasar template spectrum \citep{2016A&A...585A..87S} with the SMC extinction law \citep{2004AJ....128.1112H}. $R_V = 2.93$ is assumed \citep{1992ApJ...395..130P} (see Section 4 for discussion on the choice of extinction laws). 
The second model is the same quasar template plus one of four galaxy templates (E, Im, Sbc, or Scd, taken from \citealt{1980ApJS...43..393C}) representing the quasar host. 
The redshifts were fixed to the values in Table \ref{table:candidates}, determined with spectroscopy, in the fitting.

\begin{table*}[h]
 \caption{IR magnitudes of the red quasar candidates}
 \label{table:mag_info}
 \centering
  \begin{tabular}{lccccccc}
  \hline\hline
  Object 			& $J_{\rm AB}\rm(mag)$ & $H_{\rm AB}\rm(mag)$ & $K_{\rm AB}\rm(mag)$ & $W1_{\rm AB}\rm(mag)$ & $W2_{\rm AB}\rm(mag)$
  \\
  \hline
  J1205$-$0000  	& $21.95 \pm 0.21^{\rm v}$ & $21.48 \pm 0.34^{\rm v}$ & $20.73 \pm 0.18^{\rm v}$ & $20.68 \pm 0.18$ & $20.47 \pm 0.27$\\
  J0923$+$0402	& $20.02 \pm 0.09^{\rm u}$ & $19.74 \pm 0.16^{\rm u}$ & $19.32 \pm 0.09^{\rm u}$ & $19.42 \pm 0.08$ & $19.68 \pm 0.16$\\
  J1146$-$0154		& $> 22.15^{\rm v}$ 		   & $> 21.51^{\rm v}$ 	     & $> 21.45^{\rm v}$ 	       & $21.36 \pm 0.21$ & $20.92 \pm 0.42$\\
  J0238$-$0318		& \dots 				   & \dots 				     & \dots 			       & $20.57 \pm 0.21$ & $20.71 \pm 0.50$\\
  \hline
\end{tabular}
\begin{tabnote}
\raggedright
\textbf{Notes. }
--- The IR magnitudes used in the SED fitting. 
The near-IR magnitudes (measured in 1.0 arcsec aperture radius) were taken from the VIKING (DR5\footnote[2], labeled v) or UKIDSS (DR11\footnote[3], labeled u). 
The magnitude lower limits are placed at $5\sigma$ significance. 
The $\it WISE$ magnitudes of J0238$-$0318 were taken from the AllWISE catalog, while those of the other candidates were derived with our image decomposition analysis (see text).\\
\dag \ \url{http://horus.roe.ac.uk/vsa/index.html} \quad \ddag \ \url{http://wsa.roe.ac.uk/index.html}
\end{tabnote}
\end{table*}

\begin{figure*}[h]
 \begin{center}

       \includegraphics[clip, width=16cm]{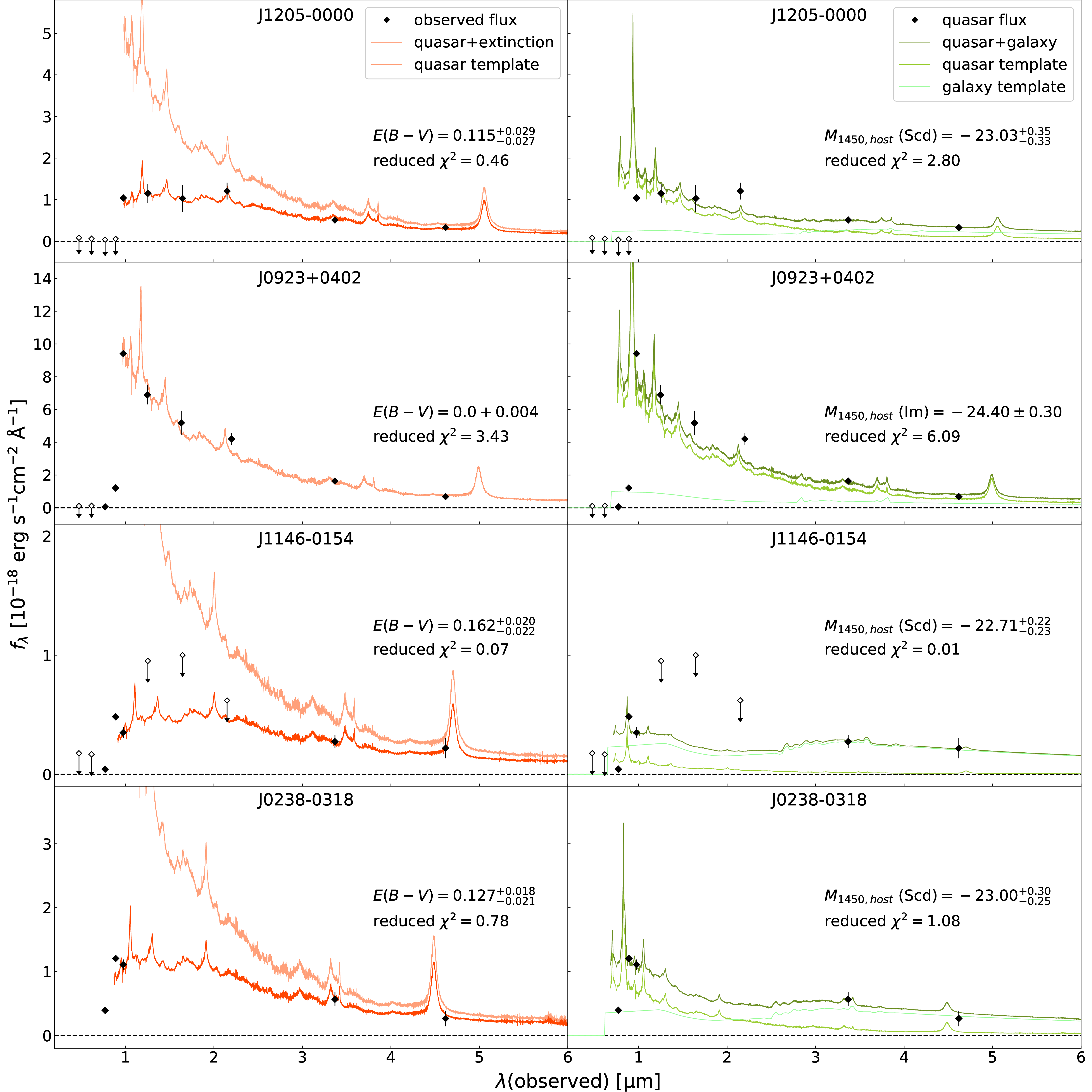} 
    
 \end{center}
\caption{The SED fitting results.
The observed fluxes and upper limits (5$\sigma$) are represented by the diamonds and arrows, respectively.
The left panels present the best-fit quasar plus extinction model, along with the quasar template before the extinction is applied.
The right panels display the best-fit quasar plus galaxy model, along with the individual quasar and galaxy templates.
The color legend is given at the top right corner of the top panels.
We report the reduced $\chi^2$, $E(B-V)$, and the host galaxy luminosity in each panel.}
\label{fig:SED_results}
\end{figure*}

Figure \ref{fig:SED_results} presents the best-fit SED and the estimated properties of each quasar. 
We present brief notes on the individual quasars in the following paragraphs. 
J1205$-$0000 and J0238$-$0318 have $E(B-V) > 0.1$ in the quasar plus dust extinction model, which gives better fits than the quasar plus galaxy model, and thus are identified as red quasars. 
J1146$-$0154 also has $E(B-V) > 0.1$, but we argue that its $\it WISE$ counterpart may be chance coincidence of a foreground source.
Table \ref{table:best-fit} lists the best-fit parameters of the four candidates.

\begin{table}[t]
 \caption{Best SED-fit parameters of the red quasar candidates}
 \label{table:best-fit}
 \centering
 \scalebox{0.78}{
  \begin{tabular}{lcccc}
  \hline\hline
  \multicolumn{1}{c}{} & \multicolumn{2}{c}{Quasar plus extinction} & \multicolumn{1}{c}{Quasar plus galaxy} 	\\
  Object 			& $E(B-V)$  & $M_{1450, \rm quasar}$\footnote[2]\ (mag)	& $M_{1450, \rm host}$ (mag) 	\\ 
  \hline
  J1205$-$0000		& $0.115^{+0.029}_{-0.027}$	& $-26.10^{+0.39}_{-0.36}$	& $-23.03^{+0.35}_{-0.33}$	\\
  J0923$+$0402	& $0.0 + 0.004$			& $-26.18^{+0.15}_{-0.14}$	& $-24.40 \pm 0.30$			\\
  J1146$-$0154		& $0.162^{+0.020}_{+0.022}$	& $-25.61^{+0.28}_{-0.30}$	& $-22.71^{+0.22}_{-0.23}$	\\	
  J0238$-$0318		& $0.127^{+0.018}_{-0.021}$	& $-25.65^{+0.24}_{-0.28}$ 	& $-23.00^{+0.30}_{-0.25}$	\\
  \hline
  
\end{tabular}
}
\begin{tabnote}
\centering
 \dag \ Intrinsic quasar luminosity corrected for the dust extinction.
\end{tabnote}
\end{table}


J1205$-$0000: this is a $z$-band dropout source and the $y$-band flux is affected by intergalactic medium absorption, so we fitted the SED models to the VIKING $J, H, Ks$-band fluxes and the decomposed $W1, W2$-band fluxes. 
The quasar plus extinction model with $E(B-V) = 0.115^{+0.029}_{-0.027}$ gives a significantly better fit than the quasar plus galaxy model. 
In addition, the host galaxy inferred in the latter case ($M_{1450} \sim -23.0$ mag) is very luminous, given the characteristic magnitude of the $z \sim 7$ galaxy luminosity function ($M^*_{1450} \sim -20.8$ mag; \citealt{2018PASJ...70S..10O}), but we found its HSC image being consistent with a point source, without apparent contribution from the host galaxy (this is however at the rest-frame $\sim$1400 \AA, and the situation can be different at 4000--6000 \AA \ traced by the $\it WISE$ bands).
We concluded that the quasar plus extinction model is more reasonable, and that this object is a red quasar.
A recently published near-IR spectrum of this quasar \citep{2019ApJ...880...77O} has a very flat continuum in rest-frame 2000--3000 \AA, consistent with our SED fitting result.
We revisited the luminosity and black hole mass ($M_{\rm BH}$) estimates given in \citet{2019ApJ...880...77O}, by taking into account the effect of dust extinction. 
As listed in Table \ref{table:BH_mass}, the extinction-corrected AGN luminosity is almost twice of the previous estimate, while the change of the $M_{\rm BH}$ and Eddington ratio  is within the measurement errors.
The near-IR spectrum shows a CIV BAL feature, but we confirmed that it doesn't have a significant impact on the $JHKs$ photometry used for the SED fitting.

\begin{table}[t]
 \caption{Luminosity and black hole mass estimates of J1205$-$0000}
 \label{table:BH_mass}
 \centering
 \scalebox{0.8}{
  \begin{tabular}{lcccc}
  \hline\hline
  	 								& \citet{2019ApJ...880...77O}	& Extinction-corrected		\\
  \hline
  $\lambda L_{3000}\ (\rm 10^{45}\ erg\ s^{-1})$		& $8.96\pm{0.66}$		& $16.15^{+2.68}_{-2.53}$	\\
  $M_{\rm BH}(\rm{Mg\ II})\ (10^9\ \it M_{\odot})$	& $2.2^{+0.2}_{-0.6}$	& $2.9^{+0.3}_{-0.8}$		\\
  $L_{\rm bol} / L_{\rm Edd}$ 					& $0.16^{+0.04}_{-0.02}$	& $0.22^{+0.04}_{-0.03}$		\\
  \hline
  
\end{tabular}
}
\end{table}

J0923$+$0402: this is an $i$-band dropout source, and we fitted the SED models to the HSC $y$-band, UKIDSS $J, H, K$-band fluxes and the decomposed $W1, W2$-band fluxes.
The SED is reproduced by a quasar template without dust extinction, so J0923$+$0402 is a normal extinction-free quasar.
This quasar is detected by $\it WISE$ simply due to its brightness; it is much brighter in the UV than the other three candidates (see Table \ref{table:candidates}).

J1146$-$0154: this is an $i$-band dropout source and is not detected in the near-IR bands (the flux upper limits are given in Figure \ref{fig:SED_results}), so we fitted the SED models to the HSC $y$-band flux and the decomposed $W1, W2$-band fluxes. 
It is reasonably fitted by both the quasar plus extinction model with $E(B-V) = 0.162^{+0.020}_{-0.022}$ and the quasar plus galaxy model. 
Thus, it could be a red quasar.
However, we noticed that the $\it WISE$ flux decomposition is uncertain for this object. 
As shown in Figure \ref{fig:contour_image}, the $W1$ flux peaks at the position of the neighbor N1, and the quasar contribution to the $W1$ flux seems minor. 
And also, the $W1$ and $W2$ images are elongated in the E-W direction, while the quasar and nearby objects (N1 and N2) are arranged in the N-S direction in the HSC image.
The decomposition may be more uncertain than reflected in the formal decomposed quasar flux errors, and so we do not include this object in the sample of red quasars in the following discussion.
As we previously mentioned, one or two of our parent sample of 93 SHELLQs quasars could be matched to a foreground $\it WISE$ source by chance, and this may be the case in J1146$-$0154.

J0238$-$0318: this is an $i$-band dropout source and is not observed in the near-IR bands. 
We fitted the SED models to the HSC $y$-band flux and the AllWISE $W1, W2$-band fluxes. 
The quasar plus extinction model with $E(B-V) = 0.127^{+0.018}_{-0.021}$ and the quasar plus galaxy model with a luminous host galaxy ($M_{1450} = -23.0$ mag) give similarly good fits. 
However, the $\chi^2$ is smaller in the former case, and the host galaxy in the latter case becomes very luminous relative to the characteristic magnitude of the galaxy luminosity function at that redshift ($M^*_{1450} \sim -20.9$ mag; \citealt{2018PASJ...70S..10O}).
The source is not extended on the HSC image, and has a typical spectrum of a high-$z$ quasar in the rest-UV.
We thus regard that the quasar plus extinction model is more reasonable, and that this object is a red quasar.

\section{Discussion}

\begin{figure}[t]
 \begin{center}
  \begin{tabular}{cc}

    \begin{minipage}{0.45\hsize}
      \begin{center}
       \includegraphics[clip, width=5.cm]{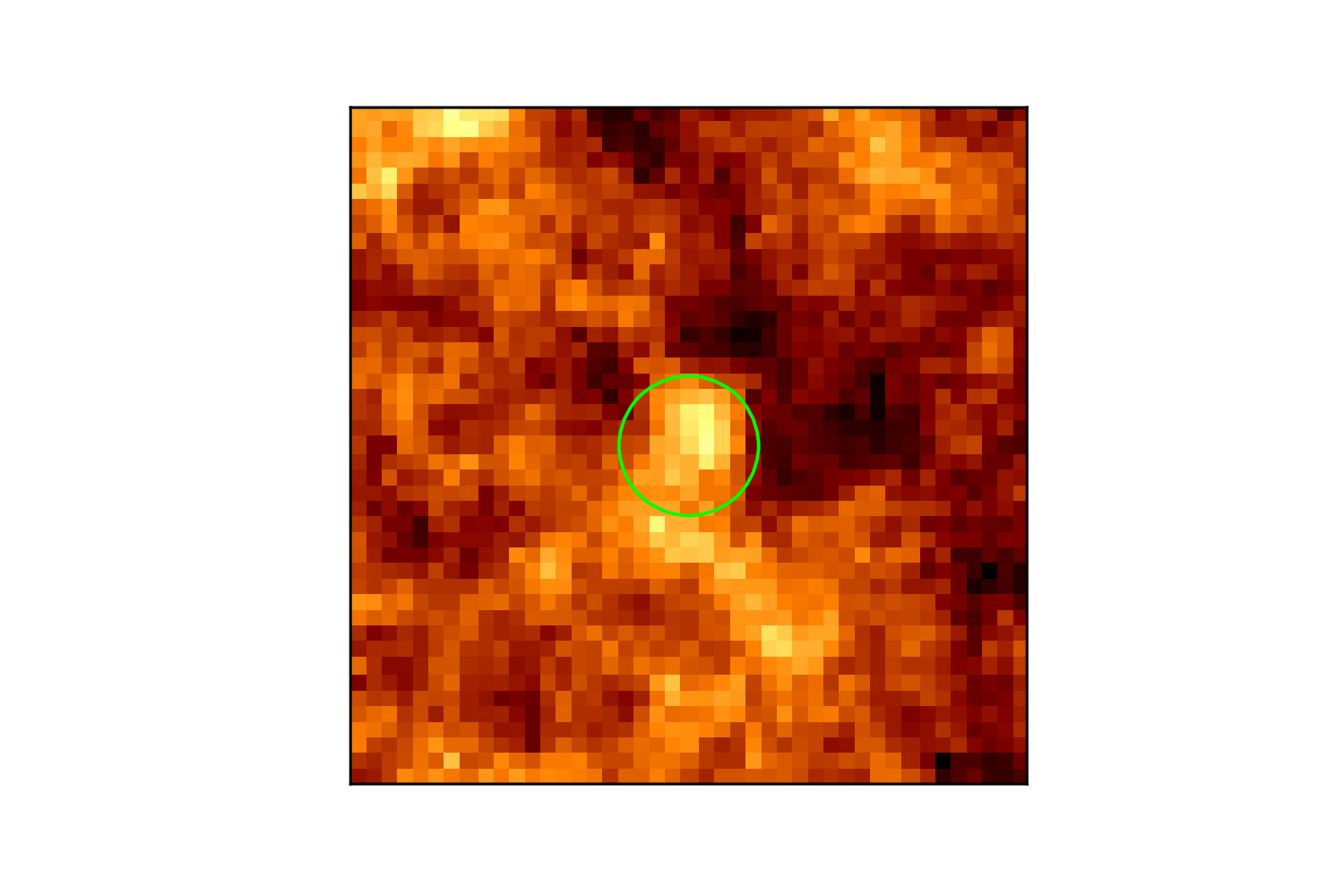} 
      \end{center}
    \end{minipage}
    
    \begin{minipage}{0.45\hsize}
      \begin{center}
       \includegraphics[clip, width=5.cm]{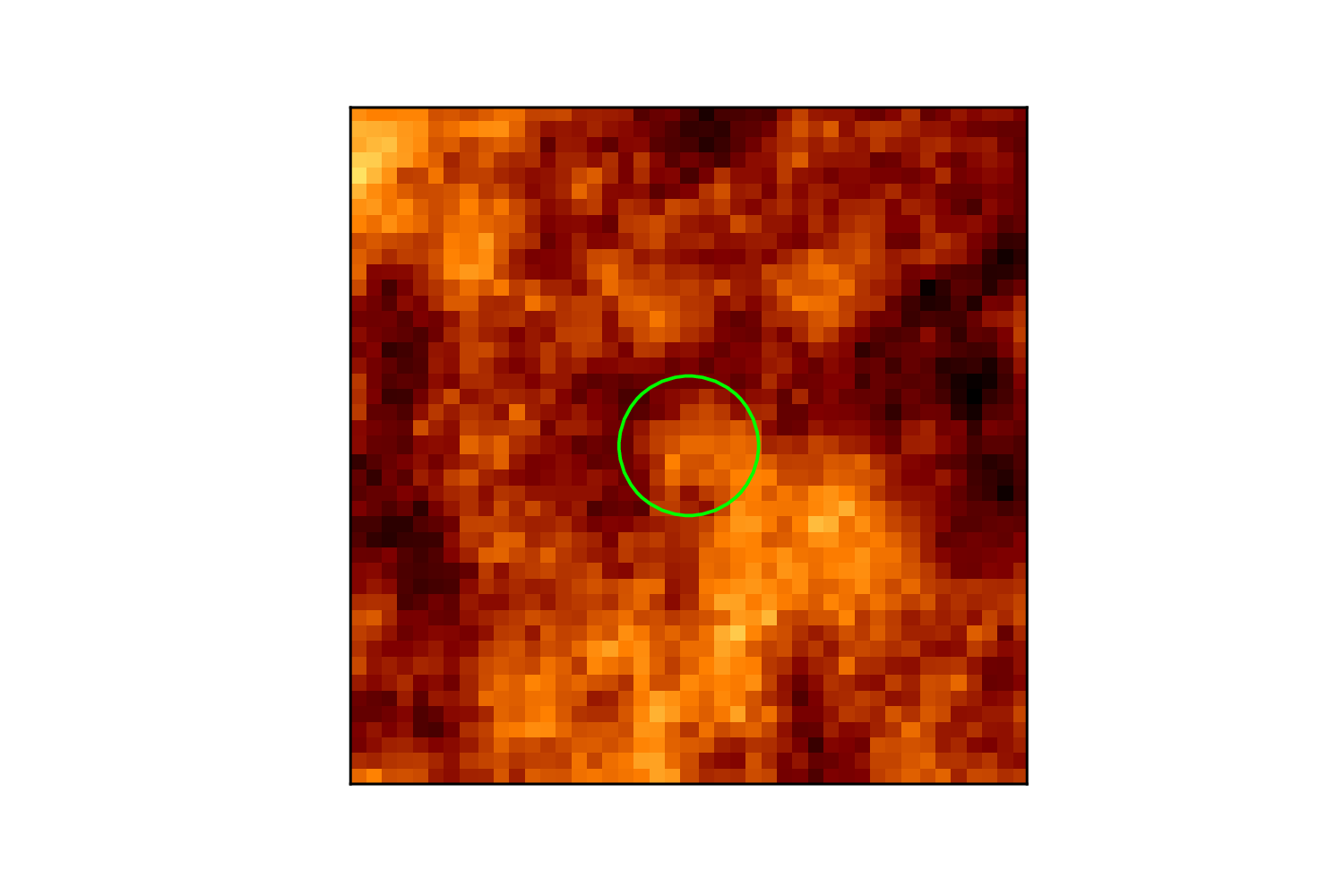} 
      \end{center}
    \end{minipage}
   
 \end{tabular}
 \end{center}
\caption{The stacked images of the 89 SHELLQs quasars without individual $\it WISE$ detection, in the $W1$ (left) and $W2$ (right) bands.
The green circles indicate the apertures (the diameter is twice the resolution, 12.2 and 12.8 arcsec in $W1$ and $W2$) used for photometry.
The image size is 60 arcsec on a side.}
\label{fig:stack_image}
\end{figure}

\begin{figure}[t]
 \begin{center}

       \includegraphics[clip, width=8cm]{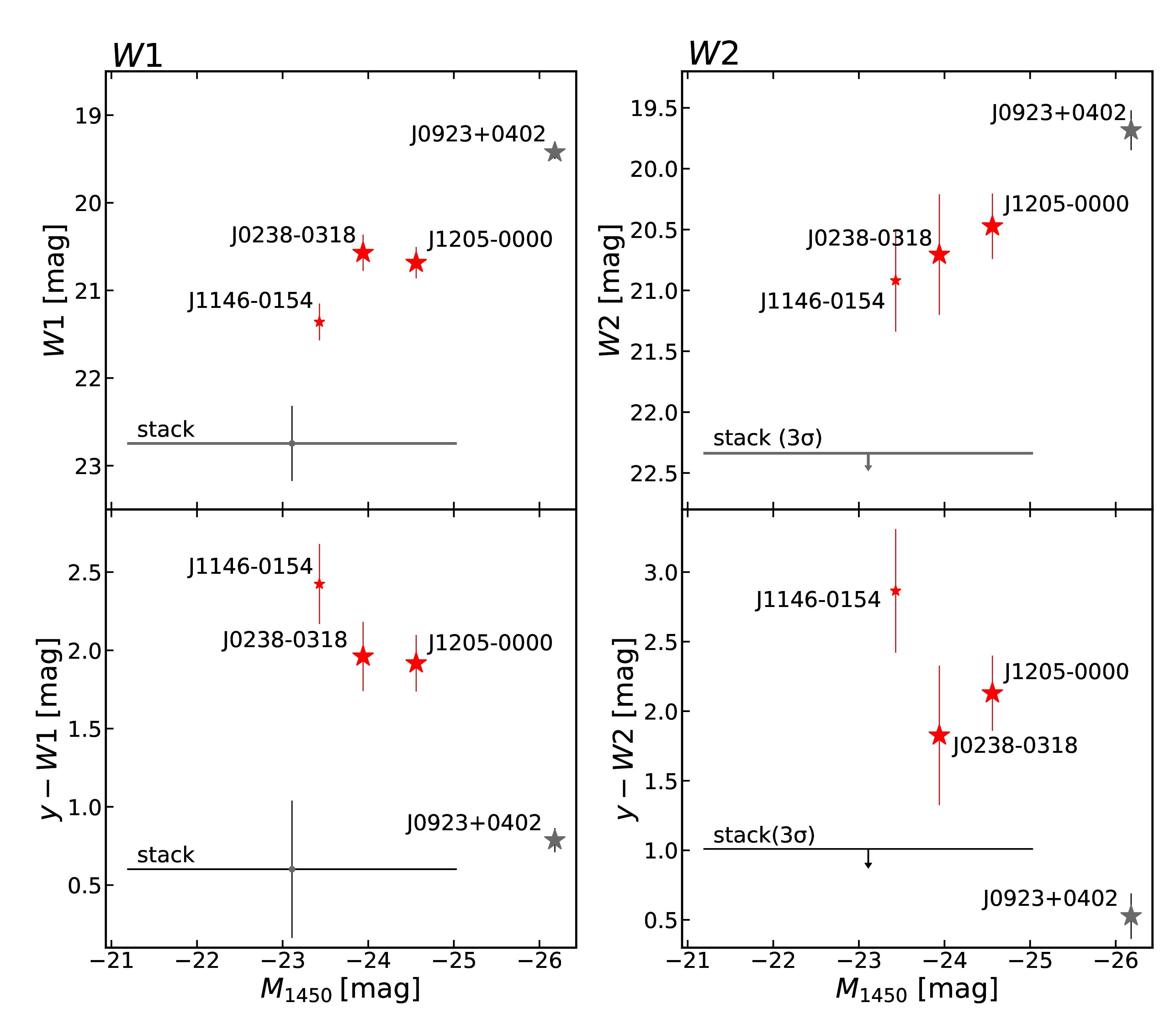} 
    
 \end{center}
\caption{The UV absolute magnitudes ($M_{\rm 1450}$) versus the $\it WISE$ magnitudes (top) and $y - \it WISE$ colors (bottom) ($W1$ on left and $W2$ on right). 
The stars represent the four SHELLQs quasars with $\it WISE$ detection (the three quasars with $E(B-V) > 0.1$ are marked in red, while the $\it WISE$ image decomposition for J1146$-$0154 (small star) is considered uncertain; see text). 
The horizontal lines represent the stacking results for the remaining 89 quasars; the arrows represent 3$\sigma$ upper limits.}
\label{fig:mag_vs_M1450}
\end{figure}

\quad Of the 93 quasars discovered by SHELLQs, only four were detected by $\it WISE$.
In order to investigate the mean properties of the remaining 89 quasars, we stacked their $\it WISE$ images centered at the HSC quasar coordinates; we did not normalize the individual images and simply calculated pixel-by-pixel median values.
We found signal ($> 3\sigma$) only in the $W1$ stacked image, as displayed in Figure \ref{fig:stack_image}. 
The aperture magnitude in the $W1$-band is $22.75 \pm 0.40$ mag, while the $3\sigma$ magnitude limit in the $W2$-band is $> 22.49$ mag.
The 89 quasars are significantly ($> 2$ mag in $W1$) fainter than the two red quasars, as shown in Figure \ref{fig:mag_vs_M1450} (top).

\begin{figure}[h]
 \begin{center}

       \includegraphics[clip, width=8cm]{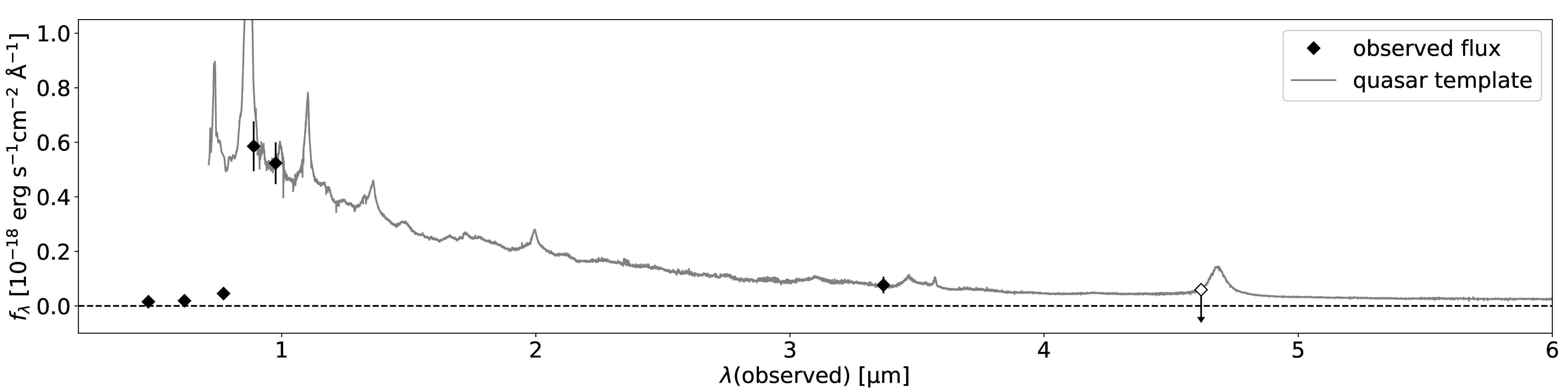} 
    
 \end{center}
\caption{Comparison of the quasar template spectrum \citep{2016A&A...585A..87S} with the mean broad-band SED of the 89 SHELLQs quasars without individual $\it WISE$ detection.
The optical data (HSC-$g, r, i, z, y$) represent the median magnitudes of the 89 quasars, while the $W1$ and $W2$ data are the result of stacking. 
The arrow in the $W2$-band represents the $3\sigma$ upper limit.
The quasar template was normalized to the observed $y$-band flux, and was drawn at the median redshift ($z = 6.13$) of the 89 quasars.}
\label{fig:SED_stack}
\end{figure}

The mean broad-band SED of the 89 quasars is presented in Figure \ref{fig:SED_stack}. 
It is well fitted by the quasar template spectrum without dust extinction (i.e., $E(B-V) = 0.0$). 
Indeed, the $y-W1$ (= 0.60) and  $y-W2$ ($< 1.0$) colors of the mean SED is significantly bluer than the two red quasars, as shown in Figure \ref{fig:mag_vs_M1450} (bottom).
The above colors are close to the colors computed from the quasar template without extinction ($y - W1 = 0.61$ and $y - W2 = 0.77$).
Thus, we concluded that the majority of the 89 quasars do not belong to the red quasar population.

For comparison with the SHELLQs low-luminosity quasars, we conducted the same analysis as described in Section 3 for more luminous quasars. 
We started from the compilation in \citet{2016ApJS..227...11B}, who listed all the high-$z$ quasars known before March 2016.
87 quasars detected by Panoramic Survey Telescope \& Rapid Response System1 (Pan-STARRS1,PS1; \citealt{2016arXiv161205560C}), and $\it WISE$ ($W1$ and $W2$ bands) were picked up among the 164 objects.
Excluding several problematic cases with very bright neighbors or irregular $\it WISE$ source shapes, we have 80 luminous ($M_{\rm 1450} < -25.4$ mag) quasars for the following analysis.

Of the 80 luminous quasars, 40 have neighbors possibly contributing to the $\it WISE$ flux, identified in the optical images.
We performed $\it WISE$ image decomposition using the approach described above.
The coordinates of the neighbors were taken from the PS1 images.
We fitted the same SED models as described above to the photometry data in the PS1 $y$-band, $J$-band and the $W1$- and $W2$-bands; the $y$- and $J$-band magnitudes were taken from \citet{2016ApJS..227...11B}, where the $J$-band magnitudes were compiled from various papers.
Four quasars were best-fitted with SEDs with $E(B-V) =$ 0.05--0.1 while all the remaining quasars have $E(B-V) < 0.05$, and thus no red quasar was found. 
In some cases we found that the quasar plus galaxy model gives a good fit, but further investigation of the possible host contribution in the luminous quasars is beyond the scope of the present analysis.

\begin{figure}[h]
 \begin{center}

       \includegraphics[clip, width=8cm]{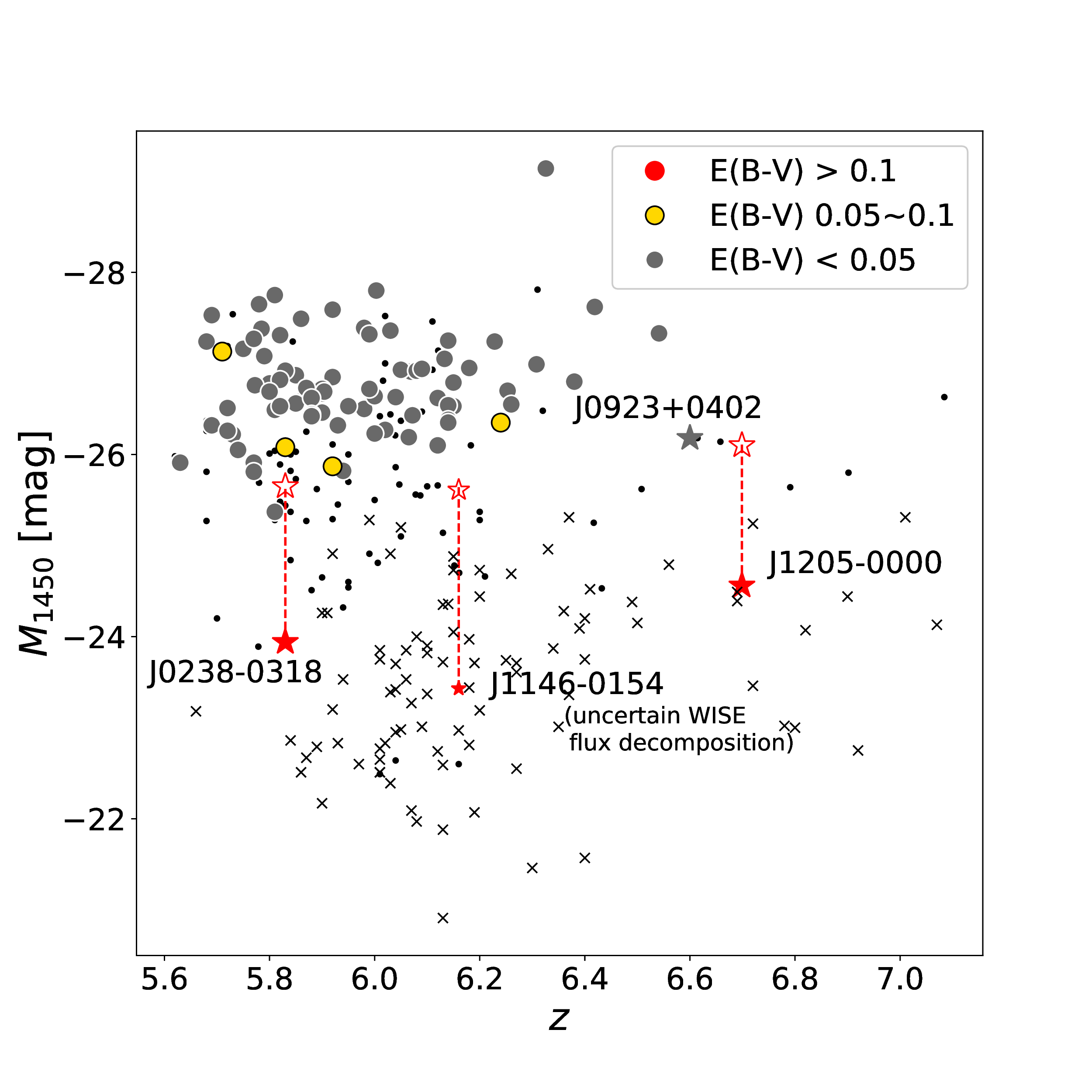} 
    
 \end{center}
\caption{The rest-UV absolute magnitudes at 1450 \AA\ ($M_{\rm 1450}$) of the four SHELLQs quasars (filled stars) and the 80 luminous quasars (filled circles) with $\it WISE$ detection, as a function of redshift. 
The colors of the points represent the amount of inferred dust extinction, as indicated at the top right corner.
The open stars represent the extinction-corrected $M_{1450}$ for the SHELLQs red quasars.
For reference, we also plot the SHELLQs quasars (crosses) and the luminous quasars in \citet[dots]{2016ApJS..227...11B} which are not analyzed in the present work, due mostly to non-detection in the $\it WISE$ bands.}
\label{fig:M1450_vs_z}
\end{figure}

Figure \ref{fig:M1450_vs_z} compares the results for the luminous sample with those for the SHELLQs quasars.
Red quasars are found only among the low-luminosity objects with $M_{1450} > -25$ mag, and the intrinsic luminosity (i.e., before affected by dust extinction) of the red quasars are around the lower envelope of the luminous quasars.
The fraction of red quasars in the SHELLQs sample is $\sim$2.2\% (2/93), and if we assume the same fraction at the luminous side, we would expect to find $\sim$4 red quasars in the 164 luminous objects. 
We found no luminous red quasar in reality, which would happen with $\sim$3\% probability based on the Poisson statistics.
So the red quasar fraction may become smaller as the observed quasar luminosity increases, but it is still inconclusive due to the small number statistics. 
Having said that, there may be two possible reasons why we do not see red quasars in the luminous sample.
A red quasar in the luminous sample would have an even brighter unextincted $M_{1450}$ than measured.
But the number density of high-$z$ quasars drops steeply above $M_{1450} \sim -25$ mag \citep{2018ApJ...869..150M}, and hence high-$z$ quasars with such brightest magnitudes are very rare.
Alternatively, it is also possible that red quasars are rarer at higher luminosity, for some physical mechanisms; for example, stronger effect of AGN feedback may quickly blow out the surrounding material, resulting in a shorter timescale of being observed as red quasars.

Overall, the observed fraction of high-$z$ red quasars seems very low.
Indeed, previous studies found the fraction at lower redshifts between $\sim$15 and $\sim$30\% \citep{2003AJ....126.1131R, 2004ApJ...607...60G, 2012ApJ...757...51G, 2018ApJ...861...37G, 2007AJ....133..186L, 2013ApJS..208...24L}.
\citet{2003AJ....126.1131R} have reported the fraction of $\simeq$15\% among $\sim$4500 quasars with $0.3 \leq z \leq 2.2$ in the Sloan Digital Sky Survey (SDSS; \citealt{2000AJ....120.1579Y}), while \citet{2012ApJ...757...51G} have estimated the fraction of $\sim$15--20\% in the radio/near-IR selected sample at $0.13 < z < 3.1$. 
However, it is not straightforward to compare the red quasar fraction at high-$z$ and low-$z$ directly, due to various systematic differences in the selection, luminosity, and definition of a red quasar, among others.
In particular, the present estimates at high-$z$ may be biased toward a lower value, because they are based on the rest-UV surveys, which may have missed a significant population of dust-extincted quasars.

In the SED fitting procedure, we adopted the SMC extinction law in accordance with the previous studies \citep{2003AJ....126.1131R, 2004ApJ...607...60G, 2012ApJ...757...51G, 2004AJ....128.1112H}, since the low metallicity of the SMC is expected to be close to the environment in high-$z$ galaxies. 
In order to assess the dependence of our results on the extinction law, we repeated the SED fitting with the extinction laws of the Large Magellanic Cloud (LMC) with $R_V = 3.16$, the Milky Way \citep{1992ApJ...395..130P} with $R_V = 3.08$, and starburst galaxies \citep{2000ApJ...533..682C} with $R_V = 3.1$.
The $E(B-V)$ values of the two red quasars are 0.117, 0.139, 0.176 and 0.321 for J1205$-$0000, and 0.127, 0.165, 0.256 and 0.458 for J0238$-$0318, with the extinction laws of SMC, LMC, Milky Way, and starbursts, respectively.
The results are consistent with the trend reported in \citet{2012ApJ...757...51G}, that the SMC extinction law gives the lowest $E(B-V)$ values.
Thus J1205$-$0000 and J0238$-$0318 are classified as red quasars regardless of the choice of extinction law.

One concern is about quasar variability, as the HSC images were taken several years after the $\it WISE$ observations.
But the time interval is actually less than a year at $z > 6$, and the expected variability at the rest-frame $\sim$1200 \AA\ (traced by the HSC $z$/$y$ band) is $\sim$0.2 mag for the luminosity of our quasars \citep{2020ApJ...894...24K}.
We confirmed that this level of variability does not affect our results, in particular the classification of J1205$-$0000 and J0238$-$0318 into red quasars.

Finally, we note that the rest-frame UV spectra of J1205$-$0000 and J1146$-$0154 have BAL features (see Figure \ref{fig:optical_spectrum}).
BAL features are thought to be generated by gas outflow ejected at high velocities from the vicinity of a quasar.
The BAL fraction among the full quasars population at $z \lesssim 4$ is estimated to be $\sim$10--15\% (e.g. \citealt{2002ApJ...578L..31T, 2003AJ....125.1784H, 2003AJ....126.2594R, 2009ApJ...692..758G}).
Some previous studies suggested the redshift dependence of the BAL quasar fraction \citep{2002ApJ...578L..31T, 2011MNRAS.410..860A}.
Indeed, \citet{2019ApJ...884...30W} found that the BAL fraction at $z \gtrsim 6.5$ is $\gtrsim 22\%$, which is a little higher than that at lower redshift.
It is known that BAL quasars, especially low-ionization BALs, are redder than general quasars \citep{1992ApJ...390...39S}.
It has been also reported that red quasars have a large BAL fraction.
\citet{2003AJ....126.1131R} found that the fraction is 20\% among 96 dust-reddened quasars in the SDSS at $1.7 \leq z \leq 2.2$.
\citet{2009ApJ...698.1095U} estimated the conservative BAL fraction of 37\% for their radio/near-IR selected 19 quasars at $z > 0.9$.
\citet{2008ApJ...672..108D} suggest that quasar samples selected in the near-IR have a larger BAL fraction than those selected in the optical.
In our two confirmed red quasars (J1205$-$0000 and J0238$-$0318), at least one has a BAL feature.
Of course this is a small sample, and a larger sample is needed to have statically meaningful results. 
Having said that, our results might indicate that high-$z$ red quasars are also preferentially associated with BAL.
The large BAL fraction would indicate that red quasars are in the dust- and gas-enshrouded phase of quasar evolution, with outflows that may eventually blow out the surrounding materials.

Since BAL may affect the broadband magnitudes used above, we repeated SED fitting by replacing the \citet{2016A&A...585A..87S} quasar template with a BAL template provided by \citet[their high-ionization BALs (HiBALs) template in Figure 5]{2019MNRAS.483.1808H}.
Since the BAL template only covers the rest-frame wavelengths $< 3000$ \AA, we kept using the \citet{2016A&A...585A..87S} template at 3000--6500 \AA, where no strong absorption is present in HiBAL quasars. 
When corrected for the intrinsic extinction of $E(B-V) \sim$ 0.023 in the BAL template \citep{2003AJ....125.1711R,2003AJ....126.2594R}, our two BAL quasars have $E(B-V) = 0.111^{+0.028}_{-0.027}$ (J1205$-$0000) and $0.158^{+0.021}_{-0.023}$ (J1146$-$0154). 
Thus J1205$-$0000 remains to be a red quasar, and our conclusions remain the same.

\section{Summary}

\quad This paper presented the discovery of two dust-reddened quasars among the 93 SHELLQs quasars.
We pre-selected four candidates based on the $\it WISE$ detection in the $W1$- and $W2$-bands, and performed SED fitting to see whether the candidates meet the red quasar definition, $E(B-V) \geq 0.1$, following \citet{2012ApJ...757...51G}.  
Of the four red quasar candidates, three have bright neighbors in the HSC images.
Because these neighbors could contribute to the AllWISE catalog fluxes, we modeled and decomposed the $\it WISE$ images as a superposition of PSFs.
We used two different models for the SED fitting; a typical quasar template with the SMC extinction law, and the same quasar template with one of four empirical galaxy templates.
As a result, we found that two quasars, J1205$-$0000 and J0238$-$0318, are red quasars.
We found that the remaining 89 SHELLQs quasars without individual $\it WISE$ detections are significantly fainter in the $\it WISE$ bands and bluer than the red quasars, and have a mean (stacked) spectrum consistent with an extinction-free quasar template.
We also carried out the same analysis for 80 luminous quasars at $z > 5.6$ taken from \citet{2016ApJS..227...11B}, but no red quasar was found.
This demonstrates the power of the wide and deep survey by HSC, which enabled us to identify high-$z$ red quasars for the first time.

The number of high-$z$ red quasars is still insufficient to statistically discuss their properties.
We will continue to search for high-$z$ red quasars, with the progress of the HSC-SSP survey.
In the short term, we plan to pursue the possibility of using the $Spitzer$ Infrared Array Camera data with better angular resolution and sensitivity than $\it WISE$ , though the analysis would be limited to the HSC Deep fields.
Future large projects such as the Rubin Observatory Legacy Survey of Space and Time in the optical and $Euclid$ in the near-IR would be useful to further advance surveys for red quasars, while deep follow-up observations with Thirty Meter Telescope, $James\ Webb\ Space\ Telescope$, and other facilities would play critical roles in studying their individual nature in detail.





\begin{ack}
We are grateful to the referee for his/her useful comments to improve this paper.
Y.M. was supported by the Japan Society for the Promotion of Science (JSPS) KAKENHI grant No. JP17H04830 and the Mitsubishi Foundation grant No. 30140.
T.I. acknowledges supports from the JSPS grant No. JP17K14247.

The Hyper Suprime-Cam (HSC) collaboration includes the astronomical communities of Japan and Taiwan, and Princeton University.  The HSC instrumentation and software were developed by the National Astronomical Observatory of Japan (NAOJ), the Kavli Institute for the Physics and Mathematics of the Universe (Kavli IPMU), the University of Tokyo, the High Energy Accelerator Research Organization (KEK), the Academia Sinica Institute for Astronomy and Astrophysics in Taiwan (ASIAA), and Princeton University.  Funding was contributed by the FIRST program from Japanese Cabinet Office, the Ministry of Education, Culture, Sports, Science and Technology (MEXT), the Japan Society for the Promotion of Science (JSPS),  Japan Science and Technology Agency  (JST),  the Toray Science  Foundation, NAOJ, Kavli IPMU, KEK, ASIAA,  and Princeton University.

The Pan-STARRS1 Surveys (PS1) have been made possible through contributions of the Institute for Astronomy, the University of Hawaii, the Pan-STARRS Project Office, the Max-Planck Society and its participating institutes, the Max Planck Institute for Astronomy, Heidelberg and the Max Planck Institute for Extraterrestrial Physics, Garching, The Johns Hopkins University, Durham University, the University of Edinburgh, Queen's University Belfast, the Harvard-Smithsonian Center for Astrophysics, the Las Cumbres Observatory Global Telescope Network Incorporated, the National Central University of Taiwan, the Space Telescope Science Institute, the National Aeronautics and Space Administration under Grant No. NNX08AR22G issued through the Planetary Science Division of the NASA Science Mission Directorate, the National Science Foundation under Grant No. AST-1238877, the University of Maryland, and Eotvos Lorand University (ELTE).
 
This paper makes use of software developed for the Large Synoptic Survey Telescope. We thank the LSST Project for making their code available as free software at http://dm.lsst.org.

This publication makes use of data products from the Wide-field Infrared Survey Explorer, which is a joint project of the University of California, Los Angeles, and the Jet Propulsion Laboratory/California Institute of Technology, funded by the National Aeronautics and Space Administration.

\end{ack}




\begin{thebibliography}{}
\bibitem[Aihara et al.(2018)]{2018PASJ...70S...4A} Aihara, H., Arimoto, N., Armstrong, R., et al.\ 2018, \pasj, 70, S4
\bibitem[Aihara et al.(2019)]{2019PASJ...71..114A} Aihara, H., AlSayyad, Y., Ando, M., et al.\ 2019, \pasj, 71, 114
\bibitem[Allen et al.(2011)]{2011MNRAS.410..860A} Allen, J.~T., Hewett, P.~C., Maddox, N., et al.\ 2011, \mnras, 410, 860
\bibitem[Arnaboldi et al.(2007)]{2007Msngr.127...28A} Arnaboldi, M., Neeser, M.~J., Parker, L.~C., et al.\ 2007, The Messenger, 127, 28
\bibitem[Ba{\~n}ados et al.(2014)]{2014AJ....148...14B} Ba{\~n}ados, E., Venemans, B.~P., Morganson, E., et al.\ 2014, \aj, 148, 14
\bibitem[Ba{\~n}ados et al.(2016)]{2016ApJS..227...11B} Ba{\~n}ados, E., Venemans, B.~P., Decarli, R., et al.\ 2016, \apjs, 227, 11
\bibitem[Ba{\~n}ados et al.(2018)]{2018ApJ...856L..25B} Ba{\~n}ados, E., Connor, T., Stern, D., et al.\ 2018, \apjl, 856, L25
\bibitem[Bosch et al.(2018)]{2018PASJ...70S...5B} Bosch, J., Armstrong, R., Bickerton, S., et al.\ 2018, \pasj, 70, S5
\bibitem[Bower et al.(2006)]{2006MNRAS.370..645B} Bower, R.~G., Benson, A.~J., Malbon, R., et al.\ 2006, \mnras, 370, 645
\bibitem[Calzetti et al.(2000)]{2000ApJ...533..682C} Calzetti, D., Armus, L., Bohlin, R.~C., et al.\ 2000, \apj, 533, 682
\bibitem[Cepa et al.(2000)]{2000SPIE.4008..623C} Cepa, J., Aguiar, M., Escalera, V.~G., et al.\ 2000, \procspie, 623
\bibitem[Chambers et al.(2016)]{2016arXiv161205560C} Chambers, K.~C., Magnier, E.~A., Metcalfe, N., et al.\ 2016, arXiv e-prints, arXiv:1612.05560
\bibitem[Ciotti \& Ostriker(2007)]{2007ApJ...665.1038C} Ciotti, L., \& Ostriker, J.~P.\ 2007, \apj, 665, 1038
\bibitem[Cole et al.(2000)]{2000MNRAS.319..168C} Cole, S., Lacey, C.~G., Baugh, C.~M., et al.\ 2000, \mnras, 319, 168
\bibitem[Coleman et al.(1980)]{1980ApJS...43..393C} Coleman, G.~D., Wu, C.-C., \& Weedman, D.~W.\ 1980, \apjs, 43, 393
\bibitem[Dai et al.(2008)]{2008ApJ...672..108D} Dai, X., Shankar, F., \& Sivakoff, G.~R.\ 2008, \apj, 672, 108
\bibitem[De Robertis et al.(1998)]{1998ApJ...496...93D} De Robertis, M.~M., Yee, H.~K.~C., \& Hayhoe, K.\ 1998, \apj, 496, 93
\bibitem[Fan et al.(2000)]{2000AJ....120.1167F} Fan, X., White, R.~L., Davis, M., et al.\ 2000, \aj, 120, 1167
\bibitem[Fan et al.(2001)]{2001AJ....122.2833F} Fan, X., Narayanan, V.~K., Lupton, R.~H., et al.\ 2001, \aj, 122, 2833
\bibitem[Fan et al.(2003)]{2003AJ....125.1649F} Fan, X., Strauss, M.~A., Schneider, D.~P., et al.\ 2003, \aj, 125, 1649
\bibitem[Fan et al.(2004)]{2004AJ....128..515F} Fan, X., Hennawi, J.~F., Richards, G.~T., et al.\ 2004, \aj, 128, 515
\bibitem[Fan et al.(2006)]{2006AJ....131.1203F} Fan, X., Strauss, M.~A., Richards, G.~T., et al.\ 2006, \aj, 131, 1203
\bibitem[Fanidakis et al.(2011)]{2011MNRAS.410...53F} Fanidakis, N., Baugh, C.~M., Benson, A.~J., et al.\ 2011, \mnras, 410, 53
\bibitem[Furusawa et al.(2018)]{2018PASJ...70S...3F} Furusawa, H., Koike, M., Takata, T., et al.\ 2018, \pasj, 70, S3
\bibitem[Gibson et al.(2009)]{2009ApJ...692..758G} Gibson, R.~R., Jiang, L., Brandt, W.~N., et al.\ 2009, \apj, 692, 758
\bibitem[Glikman et al.(2004)]{2004ApJ...607...60G} Glikman, E., Gregg, M.~D., Lacy, M., et al.\ 2004, \apj, 607, 60
\bibitem[Glikman et al.(2007)]{2007ApJ...667..673G} Glikman, E., Helfand, D.~J., White, R.~L., et al.\ 2007, \apj, 667, 673
\bibitem[Glikman et al.(2012)]{2012ApJ...757...51G} Glikman, E., Urrutia, T., Lacy, M., et al.\ 2012, \apj, 757, 51
\bibitem[Glikman et al.(2018)]{2018ApJ...861...37G} Glikman, E., Lacy, M., LaMassa, S., et al.\ 2018, \apj, 861, 37
\bibitem[Goto(2006)]{2006MNRAS.371..769G} Goto, T.\ 2006, \mnras, 371, 769
\bibitem[Hamann et al.(2017)]{2017MNRAS.464.3431H} Hamann, F., Zakamska, N.~L., Ross, N., et al.\ 2017, \mnras, 464, 3431
\bibitem[Hamann et al.(2019)]{2019MNRAS.483.1808H} Hamann, F., Herbst, H., Paris, I., et al.\ 2019, \mnras, 483, 1808
\bibitem[Hernquist \& Mihos(1995)]{1995ApJ...448...41H} Hernquist, L., \& Mihos, J.~C.\ 1995, \apj, 448, 41
\bibitem[Hewett \& Foltz(2003)]{2003AJ....125.1784H} Hewett, P.~C., \& Foltz, C.~B.\ 2003, \aj, 125, 1784
\bibitem[Hopkins et al.(2004)]{2004AJ....128.1112H} Hopkins, P.~F., Strauss, M.~A., Hall, P.~B., et al.\ 2004, \aj, 128, 1112
\bibitem[Hopkins et al.(2008)]{2008ApJS..175..356H} Hopkins, P.~F., Hernquist, L., Cox, T.~J., et al.\ 2008, \apjs, 175, 356
\bibitem[Hopkins \& Hernquist(2009)]{2009ApJ...694..599H} Hopkins, P.~F., \& Hernquist, L.\ 2009, \apj, 694, 599
\bibitem[Hopkins et al.(2014)]{2014MNRAS.445..823H} Hopkins, P.~F., Kocevski, D.~D., \& Bundy, K.\ 2014, \mnras, 445, 823
\bibitem[Izumi et al.(2018)]{2018PASJ...70...36I} Izumi, T., Onoue, M., Shirakata, H., et al.\ 2018, \pasj, 70, 36
\bibitem[Izumi et al.(2019)]{2019PASJ...71..111I} Izumi, T., Onoue, M., Matsuoka, Y., et al.\ 2019, \pasj, 71, 111
\bibitem[Jiang et al.(2016)]{2016ApJ...833..222J} Jiang, L., McGreer, I.~D., Fan, X., et al.\ 2016, \apj, 833, 222
\bibitem[Kashikawa et al.(2002)]{2002PASJ...54..819K} Kashikawa, N., Aoki, K., Asai, R., et al.\ 2002, \pasj, 54, 819
\bibitem[Kauffmann \& Haehnelt(2000)]{2000MNRAS.311..576K} Kauffmann, G., \& Haehnelt, M.\ 2000, \mnras, 311, 576
\bibitem[Kawanomoto et al.(2018)]{2018PASJ...70...66K} Kawanomoto, S., Uraguchi, F., Komiyama, Y., et al.\ 2018, \pasj, 70, 66
\bibitem[Kimura et al.(2020)]{2020ApJ...894...24K} Kimura, Y., Yamada, T., Kokubo, M., et al.\ 2020, \apj, 894, 24
\bibitem[Komiyama et al.(2018)]{2018PASJ...70S...2K} Komiyama, Y., Obuchi, Y., Nakaya, H., et al.\ 2018, \pasj, 70, S2
\bibitem[Lacy et al.(2007)]{2007AJ....133..186L} Lacy, M., Petric, A.~O., Sajina, A., et al.\ 2007, \aj, 133, 186
\bibitem[Lacy et al.(2013)]{2013ApJS..208...24L} Lacy, M., Ridgway, S.~E., Gates, E.~L., et al.\ 2013, \apjs, 208, 24
\bibitem[Lawrence et al.(2007)]{2007MNRAS.379.1599L} Lawrence, A., Warren, S.~J., Almaini, O., et al.\ 2007, \mnras, 379, 1599
\bibitem[Lagos et al.(2008)]{2008MNRAS.388..587L} Lagos, C.~D.~P., Cora, S.~A., \& Padilla, N.~D.\ 2008, \mnras, 388, 587
\bibitem[Matsuoka et al.(2016)]{2016ApJ...828...26M} Matsuoka, Y., Onoue, M., Kashikawa, N., et al.\ 2016, \apj, 828, 26
\bibitem[Matsuoka et al.(2018a)]{2018PASJ...70S..35M} Matsuoka, Y., Onoue, M., Kashikawa, N., et al.\ 2018a, \pasj, 70, S35
\bibitem[Matsuoka et al.(2018b)]{2018ApJS..237....5M} Matsuoka, Y., Iwasawa, K., Onoue, M., et al.\ 2018b, \apjs, 237, 5
\bibitem[Matsuoka et al.(2018c)]{2018ApJ...869..150M} Matsuoka, Y., Strauss, M.~A., Kashikawa, N., et al.\ 2018c, \apj, 869, 150
\bibitem[Matsuoka et al.(2019a)]{2019ApJ...872L...2M} Matsuoka, Y., Onoue, M., Kashikawa, N., et al.\ 2019a, \apj, 872, L2
\bibitem[Matsuoka et al.(2019b)]{2019ApJ...883..183M} Matsuoka, Y., Iwasawa, K., Onoue, M., et al.\ 2019b, \apj, 883, 183
\bibitem[Mazzucchelli et al.(2017)]{2017ApJ...849...91M} Mazzucchelli, C., Ba{\~n}ados, E., Venemans, B.~P., et al.\ 2017, \apj, 849, 91
\bibitem[Miyazaki et al.(2018)]{2018PASJ...70S...1M} Miyazaki, S., Komiyama, Y., Kawanomoto, S., et al.\ 2018, \pasj, 70, S1
\bibitem[Mo et al.(1998)]{1998MNRAS.295..319M} Mo, H.~J., Mao, S., \& White, S.~D.~M.\ 1998, \mnras, 295, 319
\bibitem[Mortlock et al.(2009)]{2009A&A...505...97M} Mortlock, D.~J., Patel, M., Warren, S.~J., et al.\ 2009, \aap, 505, 97
\bibitem[Mortlock et al.(2011)]{2011Natur.474..616M} Mortlock, D.~J., Warren, S.~J., Venemans, B.~P., et al.\ 2011, \nat, 474, 616
\bibitem[Oke \& Gunn(1983)]{1983ApJ...266..713O} Oke, J.~B., \& Gunn, J.~E.\ 1983, \apj, 266, 713
\bibitem[Ono et al.(2018)]{2018PASJ...70S..10O} Ono, Y., Ouchi, M., Harikane, Y., et al.\ 2018, \pasj, 70, S10
\bibitem[Onoue et al.(2019)]{2019ApJ...880...77O} Onoue, M., Kashikawa, N., Matsuoka, Y., et al.\ 2019, \apj, 880, 77
\bibitem[Pei(1992)]{1992ApJ...395..130P} Pei, Y.~C.\ 1992, \apj, 395, 130
\bibitem[Reed et al.(2015)]{2015MNRAS.454.3952R} Reed, S.~L., McMahon, R.~G., Banerji, M., et al.\ 2015, \mnras, 454, 3952
\bibitem[Reed et al.(2017)]{2017MNRAS.468.4702R} Reed, S.~L., McMahon, R.~G., Martini, P., et al.\ 2017, \mnras, 468, 4702
\bibitem[Reed et al.(2019)]{2019MNRAS.487.1874R} Reed, S.~L., Banerji, M., Becker, G.~D., et al.\ 2019, \mnras, 487, 1874
\bibitem[Reichard et al.(2003a)]{2003AJ....125.1711R} Reichard, T.~A., Richards, G.~T., Schneider, D.~P., et al.\ 2003, \aj, 125, 1711
\bibitem[Reichard et al.(2003b)]{2003AJ....126.2594R} Reichard, T.~A., Richards, G.~T., Hall, P.~B., et al.\ 2003, \aj, 126, 2594
\bibitem[Richards et al.(2003)]{2003AJ....126.1131R} Richards, G.~T., Hall, P.~B., Vanden Berk, D.~E., et al.\ 2003, \aj, 126, 1131
\bibitem[Ross et al.(2015)]{2015MNRAS.453.3932R} Ross, N.~P., Hamann, F., Zakamska, N.~L., et al.\ 2015, \mnras, 453, 3932
\bibitem[Sanders et al.(1988)]{1988ApJ...325...74S} Sanders, D.~B., Soifer, B.~T., Elias, J.~H., et al.\ 1988, \apj, 325, 74
\bibitem[Selsing et al.(2016)]{2016A&A...585A..87S} Selsing, J., Fynbo, J.~P.~U., Christensen, L., et al.\ 2016, \aap, 585, A87
\bibitem[Sprayberry \& Foltz(1992)]{1992ApJ...390...39S} Sprayberry, D., \& Foltz, C.~B.\ 1992, \apj, 390, 39
\bibitem[Tolea et al.(2002)]{2002ApJ...578L..31T} Tolea, A., Krolik, J.~H., \& Tsvetanov, Z.\ 2002, \apjl, 578, L31
\bibitem[Urrutia et al.(2009)]{2009ApJ...698.1095U} Urrutia, T., Becker, R.~H., White, R.~L., et al.\ 2009, \apj, 698, 1095
\bibitem[Venemans et al.(2012)]{2012ApJ...751L..25V} Venemans, B.~P., McMahon, R.~G., Walter, F., et al.\ 2012, \apjl, 751, L25
\bibitem[Venemans et al.(2013)]{2013ApJ...779...24V} Venemans, B.~P., Findlay, J.~R., Sutherland, W.~J., et al.\ 2013, \apj, 779, 24
\bibitem[Venemans et al.(2015)]{2015MNRAS.453.2259V} Venemans, B.~P., Verdoes Kleijn, G.~A., Mwebaze, J., et al.\ 2015, \mnras, 453, 2259
\bibitem[Venemans et al.(2016)]{2016ApJ...816...37V} Venemans, B.~P., Walter, F., Zschaechner, L., et al.\ 2016, \apj, 816, 37
\bibitem[Villforth et al.(2012)]{2012MNRAS.426..360V} Villforth, C., Sarajedini, V., \& Koekemoer, A.\ 2012, \mnras, 426, 360
\bibitem[Wang et al.(2013)]{2013ApJ...773...44W} Wang, R., Wagg, J., Carilli, C.~L., et al.\ 2013, \apj, 773, 44
\bibitem[Wang et al.(2019)]{2019ApJ...884...30W} Wang, F., Yang, J., Fan, X., et al.\ 2019, \apj, 884, 30
\bibitem[Willott et al.(2007)]{2007AJ....134.2435W} Willott, C.~J., Delorme, P., Omont, A., et al.\ 2007, \aj, 134, 2435
\bibitem[Willott et al.(2009)]{2009AJ....137.3541W} Willott, C.~J., Delorme, P., Reyl{\'e}, C., et al.\ 2009, \aj, 137, 3541
\bibitem[Willott et al.(2010)]{2010AJ....139..906W} Willott, C.~J., Delorme, P., Reyl{\'e}, C., et al.\ 2010, \aj, 139, 906
\bibitem[Wright et al.(2010)]{2010AJ....140.1868W} Wright, E.~L., Eisenhardt, P.~R.~M., Mainzer, A.~K., et al.\ 2010, \aj, 140, 1868
\bibitem[Yang et al.(2019a)]{2019ApJ...871..199Y} Yang, J., Wang, F., Fan, X., et al.\ 2019a, \apj, 871, 199
\bibitem[Yang et al.(2019b)]{2019AJ....157..236Y} Yang, J., Wang, F., Fan, X., et al.\ 2019b, \aj, 157, 236
\bibitem[York et al.(2000)]{2000AJ....120.1579Y} York, D.~G., Adelman, J., Anderson, J.~E., et al.\ 2000, \aj, 120, 1579

 

\end{thebibliography}
\end{document}